\documentclass[sigplan,nonacm]{acmart}

\newcommand{\asplossubmissionnumber}{700}
\usepackage[normalem]{ulem}

\copyrightyear{2024}
\acmYear{2024}
\setcopyright{rightsretained}
\acmConference[ASPLOS '24]{29th ACM International Conference on Architectural Support for Programming Languages and Operating Systems, Volume 2}{April 27-May 1, 2024}{La Jolla, CA, USA}
\acmBooktitle{29th ACM International Conference on Architectural Support for Programming Languages and Operating Systems, Volume 2 (ASPLOS '24), April 27-May 1, 2024, La Jolla, CA, USA}
\acmDOI{10.1145/3620665.3640428}
\acmISBN{979-8-4007-0385-0/24/04}

\usepackage{booktabs}   
\usepackage{subcaption} 
\usepackage{graphicx}

\begin{document}

\title{PIM-STM: Software Transactional Memory for Processing-In-Memory Systems}         


\author{André Lopes}
\email{andre.f.lopes@tecnico.ulisboa.pt}
\affiliation{
  \institution{INESC-ID \& Instituto Superior Técnico, Universidade de Lisboa}
  \streetaddress{Av. Rovisco Pais 1}
  \city{Lisbon}
  \country{Portugal}
  \postcode{1049-001}
}

\author{Daniel Castro}
\email{daniel.castro@ist.utl.pt}
\affiliation{%
  \institution{INESC-ID \& Instituto Superior Técnico, Universidade de Lisboa}
  \streetaddress{Av. Rovisco Pais 1}
  \city{Lisbon}
  \country{Portugal}
  \postcode{1049-001}
}

\author{Paolo Romano}
\email{romano@inesc-id.pt}
\affiliation{%
  \institution{INESC-ID \& Instituto Superior Técnico, Universidade de Lisboa}
  \streetaddress{Av. Rovisco Pais 1}
  \city{Lisbon}
  \country{Portugal}
  \postcode{1049-001}
}

\keywords{Processing-in-Memory; Transactional memory; Concurrent Systems; Concurrency control}  

\date{}

\begin{abstract}
Processing-In-Memory (PIM) is a novel approach that augments existing DRAM memory chips with lightweight logic. By allowing to offload computations to the PIM system, this architecture allows for circumventing the data-bottleneck problem that affects many modern workloads. 

This work tackles the problem of how to build efficient software implementations of the Transactional Memory  (TM) abstraction by introducing PIM-STM, a library that provides a range of diverse TM implementations  for  UPMEM, the first commercial PIM system. Via an extensive study we  assess the efficiency of alternative choices in the design space of TM algorithms on this emerging  architecture. We further quantify the impact of using different memory tiers of the UPMEM system (having different trade-offs for what concerns latency vs capacity) to store the  metadata used by  different TM implementations. Finally, we assess the gains achievable in terms of performance and memory efficiency when using PIM-STM to accelerate TM applications originally conceived for conventional CPU-based systems.
\end{abstract}

\maketitle
\pagestyle{plain}

\section{Introduction}
 
Modern workloads are becoming increasingly data-intensive, requiring to process large amounts of data, often with random access patterns~\cite{qureshi2007line}. These workloads often suffer from the known problem of the data movement bottleneck, where overall performance is degraded due to the sheer amount of data being transferred between main memory and the processor~\cite{ghose2019processing,mutlu2020modern}.

Recently, Processing-In-Memory (PIM) has gained some prominence as a way to tackle this issue~\cite{mutlu2020modern}. PIM, as the name suggests, departs from the traditional processing paradigm by performing computation directly in the memory device.
%
%
Recently, UPMEM has introduced the first commercially available PIM hardware~\cite{upmemhardware}. The UPMEM hardware achieves the goal of performing computation closer to memory by having multiple processing units, referred to as Data Processing Units (DPU) embedded in each memory module. 
Each DPU provides up to 24 hardware threads, which can communicate via a fast scratchpad memory region. Inter-DPU communication, conversely, has to be mediated by the CPU, which copies messages from the source DPU into the target(s) DPUs. 

Given that each DPU supports parallel execution of multiple hardware threads, concurrent accesses to data hosted in the same DPU needs to be synchronized in order to avoid concurrency anomalies. Synchronization of concurrent code is a long-studied problem that has traditionally been tackled using lock-based schemes. Designing efficient locking schemes, though, is a notoriously complex endeavor, as they are error-prone and vulnerable to deadlocks. Furthermore, locks can hinder composability~\cite{pankratius2011study}. These considerations led to the emergence of Transactional memory (TM)~\cite{herlihy1993transactional,herlihy2003software}. TM is a simpler and more intuitive alternative to lock-based synchronization. With TM, complexity is strongly reduced, as, by leveraging the concept of transactions, programmers only need to identify which code sections have to be executed atomically, while delegating to the TM library the task of implementing the underlying synchronization scheme. 

This work investigates, to the best of our knowledge for the first time in the literature, the problem of how to develop an efficient Software-based TM (STM) for PIM devices, by introducing PIM-STM (\S\ref{sec:pim-stm}), a library that provides a range of STM implementations for the UPMEM PIM system.  %
We leverage the  STM implementations provided by the PIM-STM library to investigate the efficiency of several choices in the design space of STM algorithms, namely the use of visible vs invisible reads, write-back vs write-through, commit-time vs encounter-time locking, and Ownership-records (ORec) vs No-Ownership records (NOrec). 
These STM implementations are specialized to cope with and take advantage of the unique hardware features of PIM architecture, in particular the existence of different memory tiers with different capacity/performance trade-offs, and the availability of atomic instructions with restricted semantics (when compared to conventional synchronization primitives available on CPUs).


By introducing the first STM implementation for PIMs, PIM-STM aims to simplify the development of  applications that needs to manage concurrent access to shared state and take full advantage of the parallelism made available by these emerging hardware architectures. In order to demonstrate the usefulness of STM in the context of PIMs and evaluate the efficiency of the alternative STM implementations provided by  PIM-STM, we develop portings for the UPMEM system of  synthetic benchmarks, concurrent data-structures as well as more complex TM benchmarks originally designed for CPUs~\cite{minh2008stamp}, namely KMeans and Labyrinth (\S\ref{sec:benchmarks}). 

Our experimental results (\S\ref{sec:single-dpu}) indicate that the most robust performance across all the evaluated workloads is achieved using a design inspired by the NOrec STM for CPU. This design opts for adopting coarse meta-data in order to alleviate the instrumentation overheads otherwise incurred by approaches that rely on fine-grained (e.g., word-level) meta-data, also known as Ownership records (Orecs). 
In fact, given the relatively low degree of parallelism provided within a single DPU in the UPMEM PIM devices, 
the additional concurrency degree achievable via the use of ORecs is, in most of the considered workloads, outweighed by the benefits (in terms of reduced overhead) provided by the NOrec design. 

However, even though the NOrec design is the most robust (performance-wise) one in our study, we also show that it can be up to 2$\times$ slower than alternative STM implementations in workloads that encompass update transactions with relatively large read phases. 
Overall, our study shows that \textit{no one-size-fits-all-solution} seems to exist that can achieve optimal performance for any workload. Thus, the choice of the STM implementations strongly depends on the workload characteristics. Our work not only provides application developers with guidelines on which workload types better fit alternative STM designs, but also makes available a library (PIM-STM) that allows developers to test the performance of alternative STM designs with their own applications via trivial configuration changes.

Finally, we conduct a study to compare the speed-ups and energy-gains attainable when porting STM-based applications to use PIM-STM (and the UPMEM system) with respect to their original CPU-based implementations (\S\ref{sec:multi-dpu}). Our study highlights, on one hand, speed-ups by up to 14.53$\times$, but, on the other hand, smaller energy gains (up to 5$\times$) and in one scenario even slightly higher energy consumption (31.5\%). These results confirm the strong performance potential of the UPMEM system, but also that its current version is not equally competitive regarding energy efficiency.

\section{Background and related Work}
\label{sec:rw}
This section provides background and discusses related work on PIM (\S\ref{sec:upmem_pim}) and on Transactional Memory (\S\ref{sec:tm}).


\subsection{Processing-in-Memory} \label{sec:upmem_pim}


\noindent\textit{Background on PIM.} Two main approaches exist for implementing PIM. Processing Using Memory (PUM) where the circuit level logic of available DRAM hardware, with little to no modifications, is used to perform computation in memory. The second approach, Processing Near Memory (PNM) places dedicated computational units near memory, which, despite being less powerful than modern CPUs, support the execution of generic application logic. Conversely, PUM is less flexible than PNM since it is limited to the copy of data, bitwise operations and simple arithmetic operations \cite{mutlu2020modern}.

\begin{figure}[t!]
    \centering
    \includegraphics[width=0.85\linewidth]{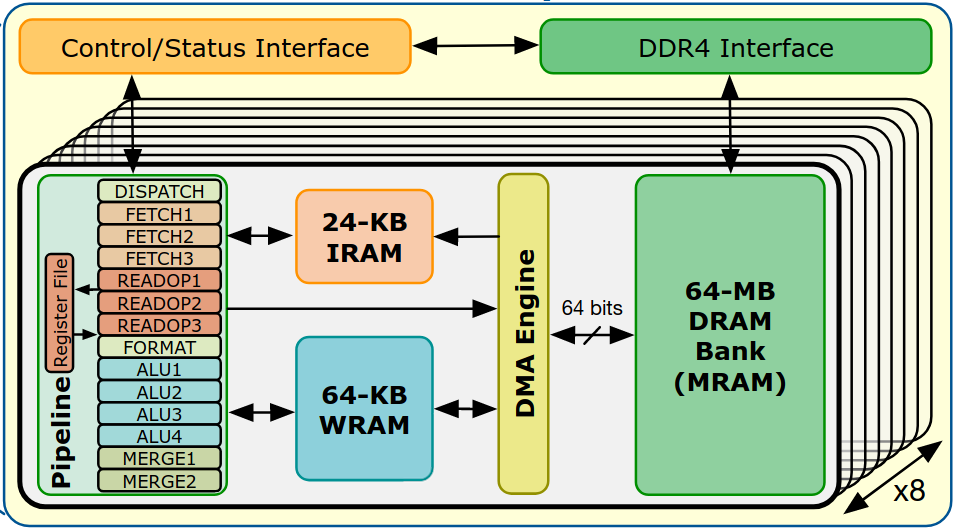}
    \caption{Internal depiction of an UPMEM PIM chip \cite{gomez2021benchmarking}.}
    \label{fig:upmem_pim}
\end{figure}

\paragraph{The UPMEM system} UPMEM PIM~\cite{gomez2021benchmarking} is the first publicly available PNM hardware implementation and it has been employed to accelerate applications in a range of domains (e.g., machine-learning~\cite{mittal2018survey,gupta2019nnpim} and bioinformatics~\cite{gupta2019rapid,diab2023framework}). An UPMEM module is a standard dual in-line memory module (DIMM) consisting of several PIM chips. A PIM chip, depicted in Fig.\ref{fig:upmem_pim} contains 8 data processing units (DPUs). Each DPU has a 64MB  DRAM bank (MRAM) that can be accessed by the CPU, 24KB of instruction memory (IRAM), 64KB of fast scratchpad memory (WRAM) and a core with 24 hardware threads. The UPMEM system has a total of 2560 DPUs, resulting in 160GB of PIM-enabled memory. 

UPMEM provides a runtime library comprised of hardware specific instructions (i.e., functions used to interact with the hardware) as well as a subset of the C standard library and a compiler (based on clang).  Using these tools, the programmer is able to write DPU programs in the C programming language. These programs use the single program multiple data (SPMD) model, where different tasklets (software threads), operating on different chunks of data, can execute different control flow paths. Each DPU can execute up to 24 tasklets concurrently (because it has 24 hardware threads), although the effective maximum degree of parallelism is achieved at 11 tasklets (parallelism is achieved by pipelining instructions from multiple tasklets via a pipeline whose maximum effective depth is 11). 

DPUs  provide two simple intra-DPU synchronization primitives, namely the acquire and release atomic instructions, which can be used to implement lock-based abstractions (shared among tasklets of the same DPU). These instructions are based on a 256 bit atomic register (i.e., a 256 bit array). The acquire/release instruction receives as input parameter a memory address and attempts to atomically acquire/release a ``logical lock'', i.e., one of the 256 entries of the bit array. The association between the input address and the index of the corresponding logical lock in the bit array is determined via a hash function implemented in hardware. Overall, the current UPMEM library provides only very simple synchronization primitives (e.g., no ReadWrite Lock implementations are provided). This renders the task of building concurrent applications for PIMs quite complex and PIM-STM aims to fill this gap by providing programmers with the abstraction of atomic transactions.


Different DPUs can execute independently  and current UPMEM systems have up to 2560 DPUs, for a total of up to 28160 concurrent tasklets. As mentioned, communication in tasklets on different DPUs has to be mediated via the CPU. In contrast, tasklets running on the same DPU can communicate via the local WRAM and MRAM. As such, to maximize the performance of the UPMEM system (and of PIM devices, more generally), applications should be engineered to increase memory access locality and avoid expensive inter-DPU communication --- a principle which is also at the basis of the design of PIM-STM, as we will discuss in \S\ref{sec:API_design}.

Another restriction of the current UPMEM system is that communication to/from the DPU can only occur when the DPU is not performing computation. More precisely, in the current UPMEM system~\cite{europar-upmem}, the CPU can access the DPU DIMMs only provided that the DPU is not performing any computation. Overall,  given that communication needs to mediated via the CPU, this limitation prevents overlapping computation and communication
--- another aspect which has affected the design of PIM-STM, as discussed in \S\ref{sec:API_design}.

\paragraph{Related work on PIM} In the literature on software libraries for PIM, the work that is most closely related to ours is SynCron \cite{giannoula2021syncron}, which implements synchronization primitives (locks, barriers, semaphores, and condition variables) across different DPUs. To implement these abstractions, Syncron assumes the availability of  message passing links that allow for efficient and direct communication among DPUs. 
A first, fundamental difference between PIM-STM and Syncron lies in the type of synchronization primitives studied, namely transactions for regulating concurrency within a DPU (PIM-STM) vs inter-dpu blocking primitives (Syncron). Further, hardware support for inter-DPU message passing is not provided by the UPMEM hardware and emulating them via CPU-mediated communications would introduce severe overheads. The absence of hardware support for direct inter-DPU communication is a key factor at the basis of PIM-STM's design choice to only support transactions that access data located in the PIM in which they execute.

PIM-STM is also related to the works that investigated the implementation of concurrent data structures for PIM~\cite{liu2017concurrent, choe2019concurrent} (which can also be implemented using STM). These works, however, assume a single threaded execution model for the DPUs and, as such, avoid concurrency issues via a simple flat-combining model~\cite{hendler2010flat}. Conversely, PIM-STM provides programmers with a generic synchronization abstraction (atomic transaction), whose implementations support the multi-threaded execution model of the UPMEM system.



\subsection{Transactional Memory}
\label{sec:tm}


Transactional memory (TM) has surfaced as a simpler and more intuitive alternative to lock-based synchronization. 
%
%
Despite borrowing the TM transaction abstraction from the database literature, STMs are not designed to operate in sand-boxed environments (unlike DBMSs). Thus, TMs normally adopt more stringent safety guarantees, such as opacity~\cite{guerraoui2008correctness}. Roughly speaking, opacity guarantees that \textit{every} transaction, including the ones that eventually abort, observes a state that can be explained via a sequential execution. Hence, opacity rules out the possibility of externalizing the writes of uncommitted transactions to concurrent transactions, a technique sometimes used in the context of database concurrency control~\cite{haritsa2000prompt,qi2023smart}. 




The TM abstraction can be implemented in software (STM), hardware (HTM) or in a combination of both (hybrid). This work investigates the problem of how to develop efficient software-based implementations of the TM abstraction for PIM devices. Therefore, in the following, we focus on analysing related works on STM implementations. STMs have been long studied in the context of cache-coherent multi-core CPUs and a plethora of alternative algorithms have been proposed in the literature, e.g.,~\cite{dalessandro2010norec,cachopo2006versioned,dragojevic2009stretching,felber2008dynamic}. In \S\ref{sec:pim-stm}, we discuss the key design choices underlying existing STM algorithms.

Recently, (S)TM implementations have been proposed for alternative types of hardware platforms, ranging from embedded devices~\cite{ferri2010embedded}, non-cache coherent many-core systems~\cite{gramoli2012tm2c}, distributed systems~\cite{bocchino2008software,couceiro2009d2stm}, GPUs~\cite{nunes2023csmv} and heterogeneous systems~\cite{castro2019hetm}. To the best of our knowledge, our work is the first to propose the use of STM for PIM and to investigate the efficiency of alternative STM designs for this emerging systems.

\section{PIM-STM} \label{sec:pim-stm}

\begin{figure}[t]
    \centering
    \includegraphics[width=\linewidth]{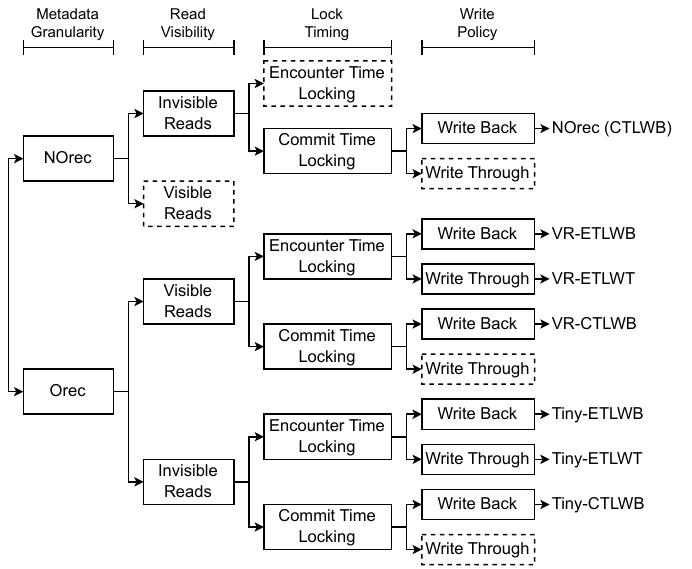}
    \caption{STM taxonomy. The designs in dashed boxes are either impossible to implement or impractical. }
    \label{fig:stm_taxonomy}
\end{figure}

This section presents the API and key design choices  (\S\ref{sec:API_design}) of the PIM-STM library, as well as the  STM implementations (\S\ref{sec:STM_algorithms}) that it includes.

\subsection{API and key design choices} \label{sec:API_design}

The PIM-STM exposes a conventional STM API that allows programs running in DPUs to demarcate (start/abort/commit) transactions and issue read/write requests to local WRAM or MRAM addresses. 
The PIM-STM library provides a number of alternative implementations of the transaction abstraction for the UPMEM PIM system and developers can control which implementation to employ via compile time macros.




A key design choice of PIM-STM is to restrict transactions to operate within the boundaries of the DPU in which they execute.
This  choice is based on two observations. First, the UPMEM system does not provide support for direct communication between DPUs and inter-DPU  communication has to be mediated by the CPU --- which is costly. In fact, we empirically verified that the latency of a CPU-mediated inter-DPU read for a 64-bit memory word is three orders of magnitude larger than a read to the local DPU MRAM (namely 331$\mu s$ vs 231 $ns$, respectively). Second, communication and computation cannot be overlapped in the current UPMEM system (\S\ref{sec:upmem_pim}), which prevents the use of speculative techniques used in the distributed TM literature to mask inter-node communication latency~\cite{peluso2012specula,Li2018HPDC}. For these reasons, we do not provide support for distributed transactions in PIM-STM and intentionally restrict their scope solely to local data --- which, as we have already mentioned, is key to maximize the performance achievable using PIM systems.

The consequence of this design decision is that programmers need to define a partitioning scheme for their applications' data that strives to avoid (or at least minimizes) the need for having to atomically manipulate data residing at different DPUs --- which in practice can still be achieved, albeit sequentially, by coordinating the data manipulation via the CPU. In fact, the complexity of defining such a  partitioning scheme is strongly application dependant, but, generally speaking, the more fine grained the data partitioning scheme, the larger are expected to be i)  the algorithmic changes required to adapt the application's logic to the new data partitioning scheme, and ii) the overheads introduced to support such algorithmic alterations (e.g., as the number of sub-problems grows, the cost for disseminating inputs and combining results grows accordingly)~\cite{palmieri-gpu}.
Overall, PIM-STM  seeks a sweet-spot regarding data partitioning by allowing developers to adopt relatively coarse data partitions (namely 64 MB, which corresponds to the memory capacity of a single DPU in the current UPMEM system), while still  providing the intuitive and familiar abstraction of transactions, although restricted to operate solely on local data.

Another key design choice is predicated on the fact that the UPMEM system has two types of memory: WRAM (fast, but with a capacity of only 64KB) and MRAM (slower, but with a capacity of 64MB). This provides the flexibility to maintain the TM metadata (such as readsets, writesets and lock table) in either of these memory modules. By placing the STM internal data structures in WRAM, the overheads of instrumentation can be reduced. However, this can, at least for some applications, create a non-trivial trade-off, as allocating the STM metadata in WRAM reduces the WRAM available for PIM applications to store their own data. The PIM-STM library controls whether STM metadata is kept in WRAM or MRAM via compile-time macros; this allows application developers to easily tune the underlying STM implementation to better match their application requirements.


\subsection{STM algorithms included in PIM-STM} \label{sec:STM_algorithms}

While designing PIM-STM, we have strived to include in it a set of implementations that could enable us to explore exhaustively the design space of STM algorithms. Given that we target programs coded using the C programming language, we focused our attention to single-version, word-based TM designs. In fact, C is a low level programming language that does not support object-orientation; further, since multi-versioning is typically used in object-oriented TMs~\cite{cachopo2006versioned}, we also dismiss this design choice. This leaves us with four main design choices, namely, meta-data granularity, visible vs invisible reads, lock timing and write policy.

\paragraph{Metadata granularity.} The granularity at which conflicts are detected is a key aspect of an STM. Two main approaches have been used in the literature, which we refer to as ORec-based vs NOrec-based. Most STM algorithms follow the ORec-based approach, where meta-data  for conflict detection are maintained in structures called Ownership Records (Orecs). This metadata is kept at the level of memory-words (or of memory regions, to limit the number of ORecs used by the STM). This design allows tracking conflicts at a finer granularity  than the alternative, NOrec design (introduced by D'Alessandro et al.~\cite{dalessandro2010norec}), which relies on a single sequence lock to track the commit event of update transactions. Whenever a transaction $T$ detects that some concurrent update transaction $T'$ committed, $T$  validates its readset, which ensures that $T'$ did not conflict with $T$. The NOrec design has the benefit of reducing the  metadata maintained and accessed during transactions' execution. Its key disadvantage is that it incurs additional validations, which are avoided by ORec based designs thanks to their ability to track conflicts at a finer granularity.


\paragraph{Read visibility.} This design choice determines whether read operations are detectable by other concurrent transactions. For this to be possible, read operations must leave a trace of their execution, which implies issuing expensive writes to shared variables. Invisible read designs avoid these costs, but cannot prevent that the value observed by an uncommitted transaction is later invalidated by a concurrently committed transaction. Therefore, approaches based on invisible reads (e.g., Tiny~\cite{felber2010time,felber2008dynamic} and NOrec~\cite{dalessandro2010norec}) rely on additional validation phases (taking place at commit time and/or during transaction execution) that verify whether the state observed by a transaction is still valid.
In cache-coherent CPUs, the visible read policy tends to perform worse~\cite{saha2006mcrt,harris2014language} due to high cache invalidation traffic that this approach generates. However, given that the architecture of the UPMEM  system is quite different from modern multi-core CPUs, we investigate whether a similar conclusion applies in this case.


\paragraph{Lock timing.} Internally, STMs regulate concurrency by using some form of locking. Regarding the time at which locks are acquired, it is possible to categorize STMs into one of the following two options: Encounter-Time Locking (ETL) and Commit-Time Locking (CTL). The former acquires locks during transaction execution, while the latter defers lock acquisition until commit time. CTL  may allow for increased concurrency by reducing the time locks are held. However, using CTL, conflicts are detected later by transactions, which can lead to performing more wasted work in case of abort. CTL also requires checking the writeset for reads-after-writes on every read. 

\paragraph{Write policy.} The write policy determines the moment in which writes are made visible. The Write-Back (WB) policy defers until commit time the task of writing new values to their memory addresses, buffering them during transaction execution. This approach avoids the cost of undoing writes when transactions abort. 
However, WB requires an additional copy phase at commit time, slowing down transactions that do not abort. The alternative approach, Write-Through (WT), writes directly to shared memory and buffers old values in an undo log, to restore old values in case the transaction aborts. With WT, reads are spared from  looking up the transaction's writeset, writes do not need to be propagated to shared memory at commit time, but it incurs the cost of undoing writes on abort.


Fig.\ref{fig:stm_taxonomy} presents the  taxonomy of the STM design choices that we consider in this work. Note that this taxonomy includes the design options that are most frequently adopted by existing STM implementations, but there are indeed some plausible, although less common, design choices that are not considered in the taxonomy (e.g., allowing transactions to wait when lock contention is encountered, rather than simply aborting). The figure highlights, using dashed boxes, the design combinations that are either incompatible (as they would break correctness) or undesirable (due to the efficiency reasons). Specifically, the WT and CTL policies are not compatible (i.e., WT is only viable with ETL), since it would lead to exposing the updates generated by uncommitted transactions (violating opacity~\cite{guerraoui2008correctness}). Further, it is undesirable to combine the NOrec design with the following two design choices (and, to the best of our knowledge, no STM algorithms exist that adopt these combinations):

\paragraph{1.~Visible reads:} as detecting the existence of concurrent transactions that read some data item, without being able to pinpoint which item was read, brings no practical advantage, thus only adding overhead.

\paragraph{2.~Encounter time locking:} as tracking via the global {sequence lock} the writes issued by ongoing transactions yields two strong disadvantages: i) it significantly amplifies the frequency of updates to the {sequence lock}, which require using atomic instructions; ii) it would lead to more frequent readset validations, which would be triggered by the write of still ongoing transactions, i.e., by transactions that can still abort in the future. Thus, these additional validations are not only expensive, but also of very little practical use. 


This leaves us with 7 viable combinations (Fig.\ref{fig:stm_taxonomy}),  corresponding to the STM algorithms included in PIM-STM, which we describe in the next section.

\subsubsection{STM implementations}

The 7 viable STM options in our taxonomy can be grouped in three main classes, namely approaches that use designs based on: i) Norec; ii) Orec and Visible reads; iii) Orec and Invisible Reads. For the first two classes, we opted for porting to UPMEM two corresponding STM algorithms for CPU, namely Tiny~\cite{felber2008dynamic,felber2010time} and NOrec~\cite{dalessandro2010norec}. We choose these algorithms as they are quite popular in the STM literature and are generally regarded as two of the most popular STM algorithms for CPU~\cite{diegues2014virtues,Diego-ProteusTM,mehrara2009parallelizing}. Algorithms based on visible reads are less common in the literature, so we developed a new STM algorithm, which we called VR. This implementation is inspired by classic lock-based concurrency control used in DBMSs and adapted to ensure STM's safety (opacity). Below we describe in more detail each of these algorithms.


\paragraph{Tiny~\cite{felber2008dynamic,felber2010time}} This implementation uses the concept of version clock validation, where a version is attributed to each memory address (and kept in an Orec alongside the lock). Every transaction maintains a lower and upper bound that constrain the visibility of its snapshot. Tiny allows extending this upper bound. An extension occurs when a transaction tries to read a memory position with a higher version than the snapshot's upper bound and requires verifying that the version of every memory address previously read has not changed since the read was performed. This extension mechanism might allow transactions from being spared from aborting, enhancing efficiency with respect to simpler designs (e.g., TL2~\cite{dice2006transactional}). To protect written positions, Tiny uses a lock table, whose entries (which serve as Orecs) cover a set of memory addresses. The mapping between an address and a lock table entry is done via a hash function. The size of the lock table (which is determined at compile time) dictates the balance between memory usage and aliasing. Aliasing happens when different memory positions are mapped to the same lock table entry. Using a larger lock table leads to less aliasing (and thus, less unnecessary aborts). However, a larger lock table also takes up more space. This is a particularly important consideration in the UPMEM hardware, given the limited capacity of WRAM. In this work, we use Tiny to fully cover the sub-tree of the taxonomy associated with Orecs and invisible reads. 


\paragraph{NOrec~\cite{dalessandro2010norec}.} This approach strives to reduce instrumentation overheads via a simple design that relies on a single sequence lock (i.e., a timestamped lock) to serialize the commit phase of update transactions. NOrec uses commit time locking and write-back as a way of decreasing the time during which transactions hold the global lock. To ensure opacity, NOrec performs value-based validation on the previously read memory locations. This is done by checking, upon every read, if any concurrent transaction committed. In the positive case, NOrec checks the read set to determine whether the values read  so far  have been overwritten by any  concurrent committed transaction. Concurrent updates are detected by verifying if the sequence lock increased. The sequence lock is also exploited to implement a simple back-off policy that delays transaction start if the lock if found busy. This helps reducing conflicts in high contention scenarios.

\begin{figure}[t]
    \centering
    \includegraphics[width=\linewidth]{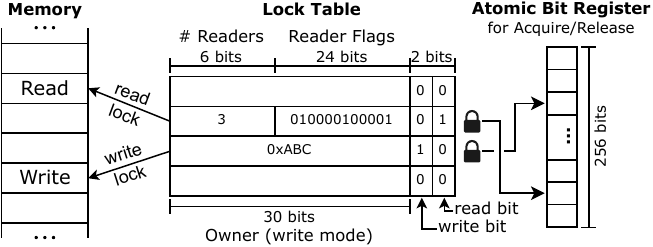}
    \caption{Design of lock table}
    \label{fig:lock_table}
\end{figure}

\paragraph{Visible Reads (VR)} This STM design tracks read and write accesses to memory words by means of read-write locks (rw-locks). Similarly to Tiny, VR use  a lock table where each entry is mapped to a set of memory addresses via  hashing. Each entry of the lock table contains a rw-lock that controls access to the corresponding memory position. This implementation contains 3 variants ({\sf VR CTLWB}, {\sf VR ETLWB} and {\sf VR ETLWT}). All these variants ensure read visibility by having transactions acquire rw-locks in read mode as soon as the read is performed. Write operations trigger the acquisition of the corresponding rw-lock in write mode, which it takes immediately or at commit time depending on the lock timing  policy (ETL vs CTL). Fig.\ref{fig:lock_table} illustrates our rw-lock implementation for the UPMEM system. Each rw-lock uses a 32-bit word. The 2 least significant bits are used to encode whether the lock is acquired and in which mode. 
If  the lock is acquired in read mode, we use the topmost 6 bits to store the number of readers currently holding the lock. As UPMEM supports at most 24 concurrent tasklets, we use the  remaining 24 bits to also store the identity of the readers that have acquired the lock. This is useful in combination with the WB policy, as it spares from having to consult the writeset whenever a read is issued (to return the latest value written by the transaction, if any). For efficiency reasons, when the lock is in write mode, we encode the lock owner identity by storing in topmost 30 bits the (word-aligned) address of the owner's readset. To avoid deadlocks, a transaction aborts every time it tries to acquire a lock that is already being held in an incompatible mode. This means that a transaction request to upgrade a read lock to a write lock causes an abort if the lock is held in read mode by other transactions.

Unlike all other approaches, the VR design avoids the need for validating  previously read memory positions. However, it incurs additional costs due to the need of tracking readers (by acquiring a read-write lock in read-mode). Further, its lock-based design makes it more susceptible to spurious aborts in high contention workloads.


\paragraph{Hardware synchronization primitives.} Existing CPU-based STM implementations, such as TinySTM and NOrec, rely heavily on compare-and-swap (CAS) to, e.g.,  update the sequence lock (NOrec) or to update an entry of the lock-table (Tiny). However, the CAS instruction is not available on the UPMEM hardware. To implement the CAS primitive (on the UPMEM hardware), we use the acquire and release instructions described in \S\ref{sec:upmem_pim}. More precisely, we first acquire a lock on the address targeted by the CAS operation, then we check if the current value matches the expected one and finally, we release the lock. Recall that the acquire and release atomic instruction are implemented via a 256 bit atomic register. Thus, when two tasklets try to acquire locks on different addresses (e.g., corresponding to different lock table entries) that are mapped to the same bit of the atomic register, the two tasklets may suffer lock aliasing and be unnecessarily serialized. However, in our STM implementations, this serialization occurs only for the time needed to consult and possibly update a lock table entry (Tiny and VR) or to update the sequence lock (NOrec). This is a relatively short period of time compared to the actual transaction duration and, as we will see in \S\ref{sec:single-dpu}, the impact of this lock aliasing on performance is negligible. Further, as the  acquire/release primitives of UPMEM operate on a hardware register (i.e., they do not access WRAM or MRAM), their overhead is minimal in practice.

\section{Experimental Evaluation}
\label{sec:exp-eval}

This section aims to answer the following key questions:
\begin{enumerate}
    \item Which STM designs perform better in different settings? What guidelines can be provided to identify which STMs better fit  different workload characteristics? (\S\ref{sec:single-dpu})
    \item What  impact has the choice of allocating the STM metadata in WRAM (vs MRAM) on the efficiency of the various STM designs? (\S\ref{sec:mram} and \S\ref{sec:wram})
    \item What performance and energy gains can be achieved by using PIM-STM to accelerate STM applications originally designed for CPUs? (See \S\ref{sec:multi-dpu})
\end{enumerate}

In order to answer the first two questions above, we  study the performance of the different STM implementations provided by PIM-STM on a single DPU. This allows us to assess the scalability and efficiency of the various solutions as the degree of parallelism on an individual DPU varies. We evaluate the scenario in which multiple DPUs are used concurrently when addressing the third of the above questions. 



All the experiments presented in this section were run on the UPMEM server, which is equipped wit two Intel Xeon Silver 4215 CPUs. This system has 256GB of main memory (DRAM) and 160GB of PIM-enabled memory (i.e., a total 2560 DPUs, see \S\ref{sec:upmem_pim}). For our multi-DPU study (\S\ref{sec:multi-dpu} study, we use a  machine equipped with an Intel Xeon Gold 5218 CPU (32  hardware threads) and 190GB of DRAM. Unless otherwise specified, the reported results are obtained by averaging 10 runs and we also report the corresponding standard deviation.

The code of PIM-STM and all the presented benchmarks is publicly available~\footnote{\url{https://github.com/Andre12Lopes/PIM-STM.git}}.



\subsection{Benchmarks} \label{sec:benchmarks}
Below we describe the set of benchmarks used in our study, which includes synthetic benchmarks, concurrent data structures as well as two complex applications from the STAMP benchmark {suite}~\cite{minh2008stamp} in the domain of machine learning (Kmeans) and VLSI design (Labyrinth).

\paragraph{ArrayBench.} This is a synthetic benchmark that relies on transactions to manipulate an array of size $N$ and that we use to shape two  workloads (denoted as A and B) with diverse characteristics. 
In workload A, $N$ is set to 12,500 and the array is split into two regions, one of size $Y$=2,500 and another of size $K$=10,000 (such that $Y$+$K$=$N$). Transactions execute in two phases: in the first phase, they read 100 random array entries from region $Y$; in the second phase, 20 array entries at random in region $K$ are read and modified. 
In workload B, we set $K$=10 and execute only the second phase, in which transactions manipulate 4 array entries. Overall, workload A is less contention prone, despite generating a larger number of read and write accesses (as the first phase operates on a non-contended array region and the array region manipulated in the second phase is much smaller in workload B).

\paragraph{Linked-List.} An implementation of a concurrent Linked-List that uses TM for synchronization. It exposes three operations: \textit{add}, \textit{remove} and \textit{contains}. Each is encapsulated within a transaction. 
The size of the list is kept roughly constant by enforcing equal number of adds and removes throughout the duration of the benchmark. 
We consider two workloads which generate different contention levels: in the low contention ({\sf LC}) workload, 90\% of the operations are \textit{contains} (i.e., read-only transactions); in the high contention ({\sf HC}) workload,  only 50\% of the operations are \textit{contains}. Each tasklet performs 100 operations and initially the list has 10 elements.

\paragraph{KMeans.} This is a TM based porting of the  K-means algorithm, whose goal is to determine the coordinates of the centroids  of $k$ clusters, given as input a set of $N$-dimensional points. This is achieved by initializing the clusters' coordinates at random and assigning each input point to the currently closest cluster and updating its centroid. 
In KMeans, transactions are used to update the coordinates of the centroid to which an input is assigned to, but the computation of the closest centroid is performed non-transactionally. Thus, transactions are relatively small (their readset and writeset size coincides with $N$) and the fraction of time spent in transactions decreases quickly as the number of centroids grows. Also in this case, we consider a low contention ({\sf LC}) workload ($k=15$, $N=14$), and a high contention ({\sf HC}) scenario ($k=2$, $N=14$).




\paragraph{Labyrinth.} The Labyrinth benchmark~\cite{minh2008stamp} is a porting of the Lee  algorithm~\cite{lee1961algorithm}. Transactions are used to concurrently route  paths over a shared 3-dimensional grid while guaranteeing that paths do not overlap. Transactions  encompass an expensive computation aimed at identifying the shortest path. However, this phase operates on a private copy of the grid that is accessed directly, i.e., without using the STM API. We use Labyrinth to generate 3 workloads that route 100 paths over grids of different sizes, namely 16$\times$16$\times$3, 32$\times$32$\times$3 and 128$\times$128$\times$3 for workloads, named S, M, and L resp. By varying the grid size, the duration of transactions (and the size of their readset/writeset) {increases accordingly}.



\subsection{Efficiency of alternative STM designs} \label{sec:single-dpu}

This study assesses the   efficiency of the alternative STM designs of PIM-STM on a single DPU, both for  the case of STM metadata kept in MRAM (\S\ref{sec:mram}) and  in WRAM (\S\ref{sec:wram}).


\subsubsection{STM metadata hosted in MRAM} \label{sec:mram}

\begin{figure*}[t!]
    \subfloat[Throughput (ArrayBench {\sf A})\label{fig:plot_throughput_bank_10000_acc_mram}
    ]{
    \includegraphics[width=0.24\linewidth]{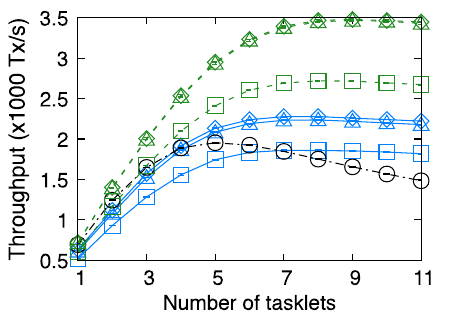}
    }
    \subfloat[Throughput (ArrayBench {\sf B})\label{fig:plot_throughput_bank_high_con_mram}
    ]{
        \includegraphics[width=0.24\linewidth]{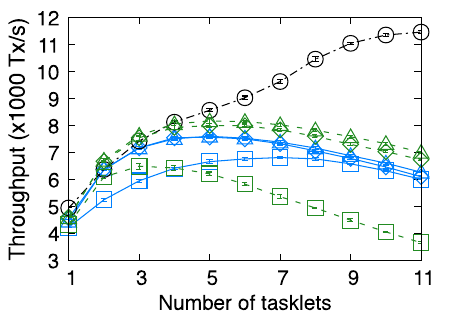}
    }
    \hfill
    \subfloat[Throughput (Linked-List {\sf LC})\label{fig:plot_throughput_linked_list_10_upd_mram}
    ]{
        \includegraphics[width=0.24\linewidth]{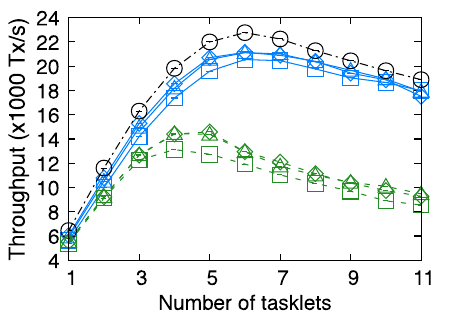}
    }
    \subfloat[Throughput (Linked-List {\sf HC})\label{fig:plot_throughput_linked_list_50_upd_mram}
    ]{
        \includegraphics[width=0.24\linewidth]{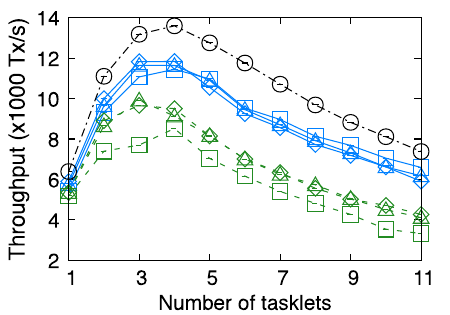}
    }
    \vspace{-2pt}
    \medskip
    \subfloat[Abort rate (ArrayBench {\sf A})\label{fig:plot_aborts_bank_10000_acc_mram}
    ]{
        \includegraphics[width=0.24\linewidth]{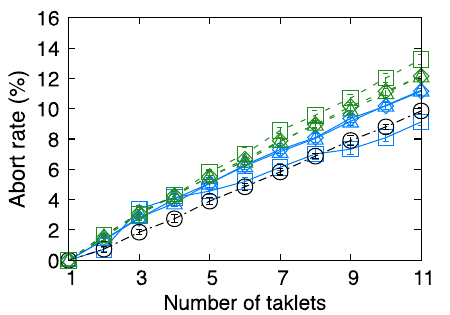}
    }
    \subfloat[Abort rate (ArrayBench {\sf B})\label{fig:plot_aborts_bank_high_con_mram}
    ]{
        \includegraphics[width=0.24\linewidth]{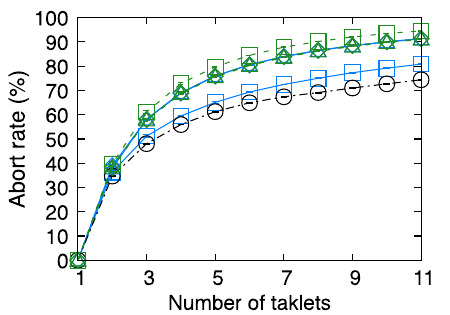}
        %
    } 
    \hfill
    \subfloat[Abort rate (Linked-List {\sf LC})\label{fig:plot_aborts_linked_list_10_upd_mram}
    ]{
        \includegraphics[width=0.24\linewidth]{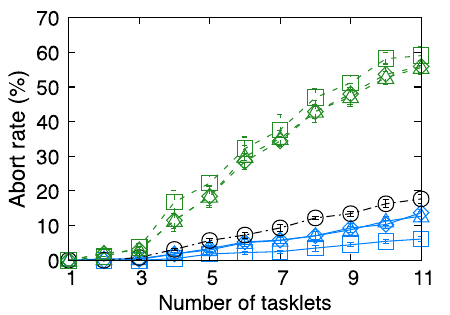}
        %
    }
    \subfloat[Abort rate (Linked-List {\sf HC})\label{fig:plot_aborts_linked_list_50_upd_mram}
    ]{
        \includegraphics[width=0.24\linewidth]{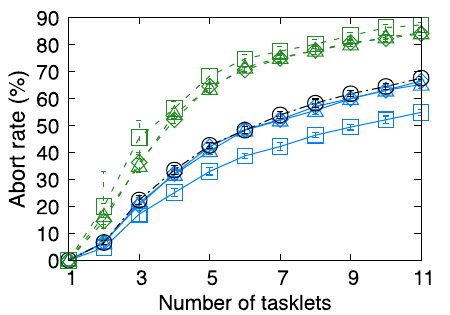}
    }
    \hfill
    \medskip   
    \subfloat[Phases (ArrayBench {\sf A})\label{fig:plot_phases_bank_bank_10000_acc_mram}]{
        \includegraphics[width=0.24\linewidth]{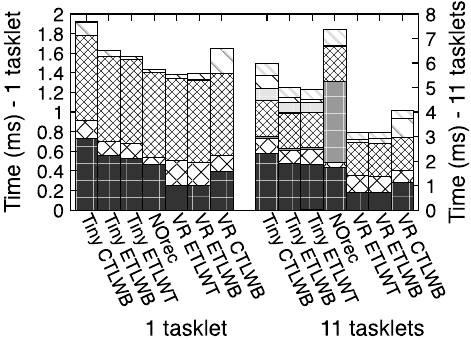}
    }
    \subfloat[Phases (ArrayBench {\sf B})\label{fig:plot_phases_bank_high_con_mram}]{
        \includegraphics[width=0.24\linewidth]{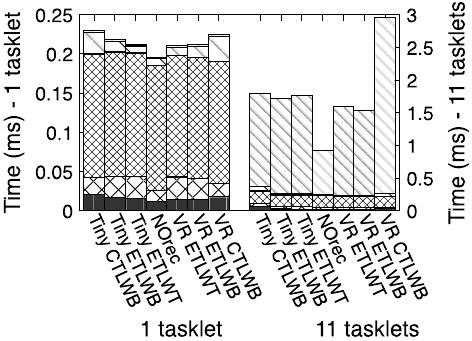}
    }
    \hfill
    \subfloat[Phases (Linked-List {\sf LC})\label{fig:plot_phases_linked_list_10_upd_mram}]{
        \includegraphics[width=0.24\linewidth]{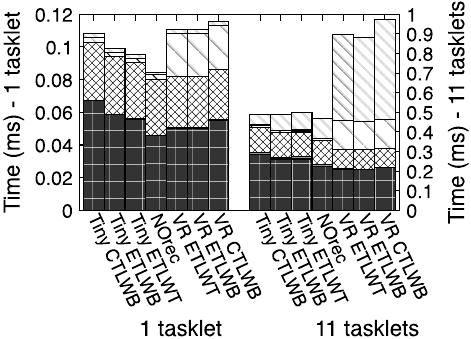}
        %
    }
    \subfloat[Phases (Linked-List {\sf HC})\label{fig:plot_phases_linked_list_50_upd_mram}]{
        \includegraphics[width=0.24\linewidth]{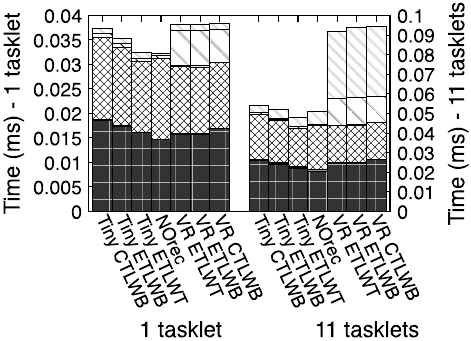}
        %
    }
    \hfill
    \medskip
    \hfill
    \begin{tabular}{c c c c c}
         \includegraphics[width=0.13\linewidth]{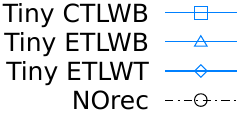} & \includegraphics[width=0.12\linewidth]{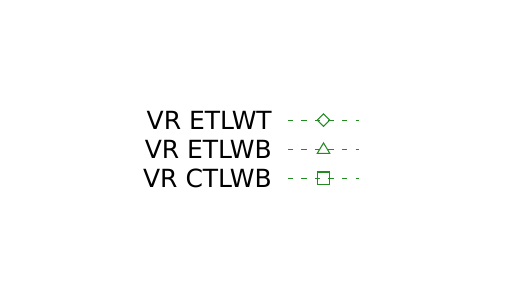} & $\qquad\qquad\qquad$ & \includegraphics[width=0.19\linewidth]{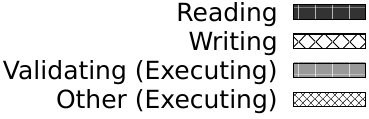} & \includegraphics[width=0.17\linewidth]{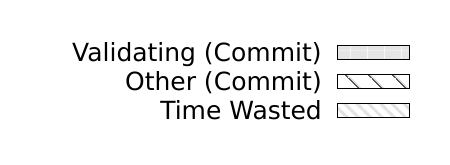} \\
    \end{tabular}
    \hfill
    \caption{Throughput, abort rate and time breakdown for ArrayBench and Linked-List with metadata in MRAM.}
    \label{fig:bank_ll_mram}
\end{figure*}

\begin{figure*}[t!]
    \subfloat[Throughput (KMeans {\sf LC})\label{fig:plot_throughput_kmeans_14_dims_15_centers_mram}
    ]{
        \includegraphics[width=0.24\linewidth]{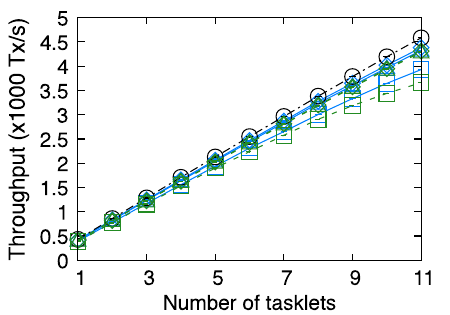}
    }
    \subfloat[Throughput (KMeans {\sf HC})\label{fig:plot_throughput_kmeans_14_dims_2_centers_mram}
    ]{
        \includegraphics[width=0.24\linewidth]  {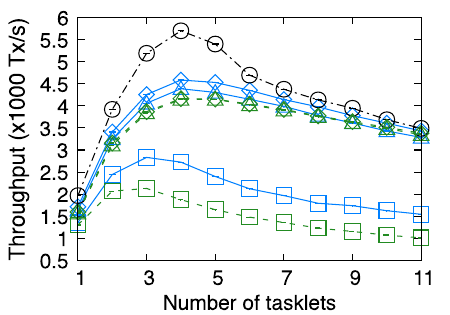}
    }
    \hfill
    \subfloat[Throughput (Labyrinth {\sf S})\label{fig:plot_throughput_labyrinth_low_con_mram}
    ]{
        \includegraphics[width=0.24\linewidth]{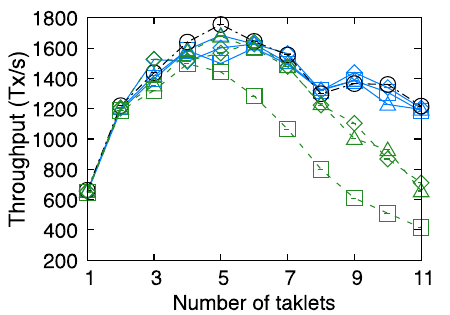}
    }
    \subfloat[Throughput (Labyrinth {\sf L})\label{fig:plot_throughput_labyrinth_high_con_mram}
    ]{
        \includegraphics[width=0.24\linewidth]{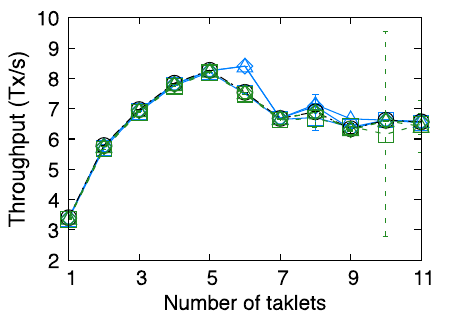}
    }
    
    \medskip
    \subfloat[Abort rate (KMeans {\sf LC})\label{fig:plot_aborts_kmeans_14_dims_15_centers_mram}
    ]{
        \includegraphics[width=0.24\linewidth]{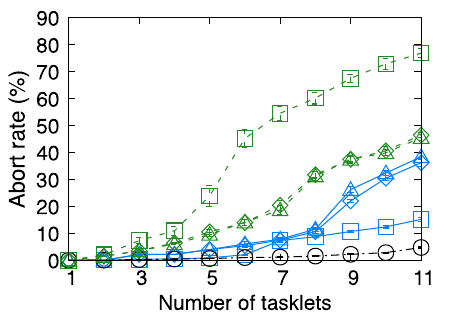}
    }
    \subfloat[Abort rate (KMeans {\sf HC})\label{fig:plot_aborts_kmeans_14_dims_2_centers_mram}
    ]{
        \includegraphics[width=0.24\linewidth]{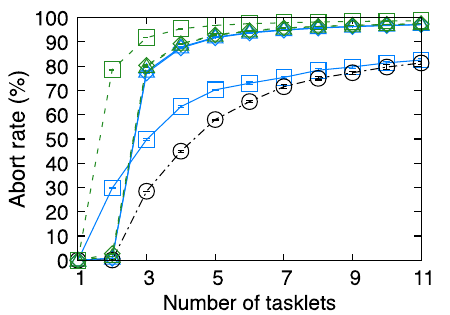}
    }
    \hfill
    \subfloat[Abort rate (Labyrinth {\sf S})\label{fig:plot_aborts_labyrinth_low_con_mram}
    ]{
        \includegraphics[width=0.24\linewidth]{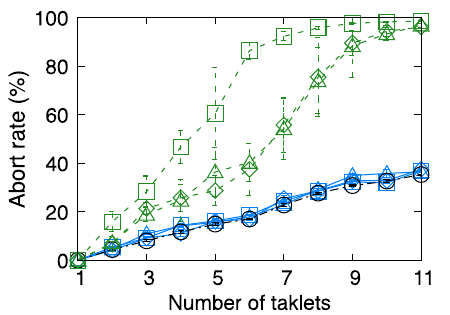}
    }
    \subfloat[Abort rate (Labyrinth {\sf L})\label{fig:plot_aborts_labyrinth_high_con_mram}
    ]{
        \includegraphics[width=0.24\linewidth]{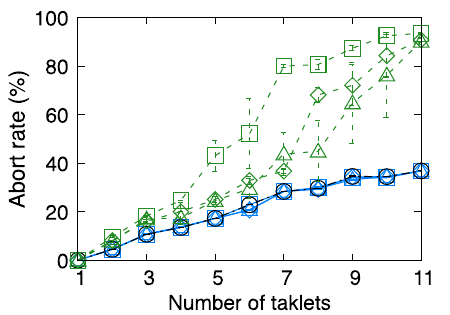}
    }
    
    \medskip   
    \subfloat[Phases (KMeans {\sf LC})\label{fig:plot_phases_kmeans_14_dims_15_centers_mram}]{
        \includegraphics[width=0.24\linewidth]{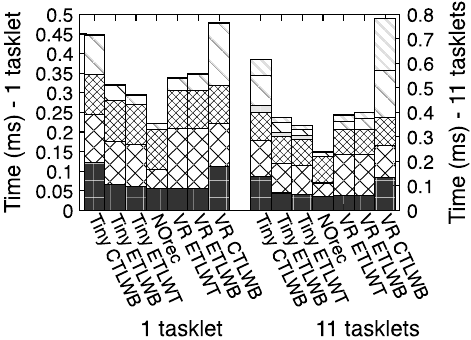}
    }
    \subfloat[Phases (KMeans {\sf HC})\label{fig:plot_phases_kmeans_14_dims_2_centers_mram}]{
        \includegraphics[width=0.24\linewidth]{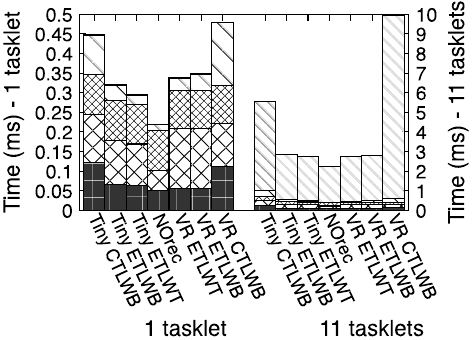}
    }
    \hfill
    \subfloat[Phases (Labyrinth {\sf S})\label{fig:plot_phases_labyrinth_16_16_3_mram}]{
        \includegraphics[width=0.24\linewidth]{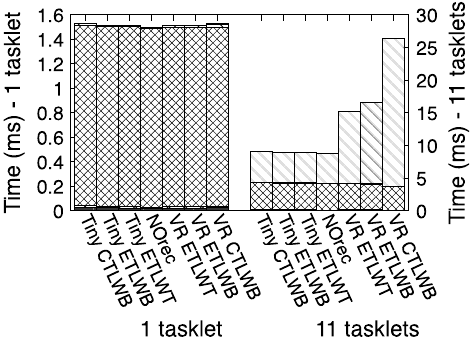}
    }
    \subfloat[Phases (Labyrinth {\sf L})\label{fig:plot_phases_labyrinth_128_128_3_mram}]{
        \includegraphics[width=0.24\linewidth]{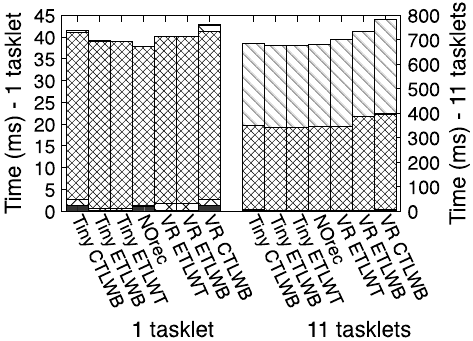}
    }
    \hfill
    \medskip
    \hfill
    \begin{tabular}{c c c c c}
         \includegraphics[width=0.13\linewidth]{Graphs/legend_1.pdf} & \includegraphics[width=0.12\linewidth]{Graphs/legend_2.pdf} & $\qquad\qquad\qquad$ & \includegraphics[width=0.19\linewidth]{Graphs/legend_phases_1.pdf} & \includegraphics[width=0.17\linewidth]{Graphs/legend_phases_2.pdf} \\
    \end{tabular}
    \hfill
    \caption{Throughput, abort rate and time breakdown for the KMeans and Labyrinth benchmark with metadata in MRAM.}
    \label{fig:kmeans_labyrinth_mram}
\end{figure*}

Fig.~\ref{fig:bank_ll_mram} and~\ref{fig:kmeans_labyrinth_mram} report the throughput (number of committed transactions per second), abort rate and time breakdown for the case where TM metadata is kept in MRAM. In the following, we analyse each benchmark. 

%





\paragraph{ArrayBench.} 
The two workloads of this benchmark have very different characteristics and this is reflected into the relative performance of the STM algorithms to the extent that the worst performing solution for workload {\sf A} is the most competitive one for workload {\sf B}.

Let us start by discussing the results for ArrayBench {\sf A}. In this case, the top 3 performing solutions are based on the VR design, with the two ETL variants delivering the best performances, followed by the CTL variant. While the choice of {the} write policy (WB vs WT) appears to have limited impact, the  CTL design reduces peak throughput by nearly 25\% with VR. This can be explained via the breakdown plot in  Fig.\ref{fig:plot_phases_bank_bank_10000_acc_mram}, which shows that {\sf VR CTLWB} spends more time in the read and commit phases than the ETL variants. 
This can be explained by considering that Tiny's CTL implementation requires scanning the write set every time a read is performed (to return values previously written by the current transaction). Additionally, all CTL implementations perform more work at commit time (i.e., acquire locks and write values to memory).


As for the invisible reads solutions, the variants based on Orecs (i.e., Tiny-based) are approx.~two times slower than the best performing VR-based variant. NOrec is the worse performing solution with this workload and at 11 tasklets it is around 2.5$\times$ slower than the best STM. By the breakdown plot in Fig.\ref{fig:plot_phases_bank_bank_10000_acc_mram}, we  see that the performance of these solutions is hampered by their additional readset validation(s), which are expensive in this workload as readsets are relatively large (transactions read up to 120  memory positions). This cost is exacerbated in NOrec, which needs to validate the transaction's readset upon each read if any update transaction commits concurrently. Thanks to its finer grained conflict detection capability, Tiny performs a lower number of validations than NOrec.
 We also see that the cost of executing read operations is larger for the solutions that use the invisible reads policy (Tiny and NOrec). This may appear counter-intuitive, given that the VR policy requires acquiring a rw-lock in read mode  (a cost that is spared by both Tiny and NOrec). However, keeping in mind that the read latency is dominated by the number of needed MRAM accesses (which overshadow the latency of WRAM and register accesses), there are two main reasons that justify these results: i) the acquisition of the rw-lock relies on an atomic operation (\S\ref{sec:upmem_pim}) that operates on a DPU register, i.e., it does not access MRAM and, thus, introduces very limited overhead; ii) the Tiny and NOrec variants execute a larger number of MRAM read accesses than the VR variants, some of these accesses being  due to the invisible read design (e.g., reading twice the lock to detect concurrent writes or reading the  transaction snapshot) and others (for the case of NOrec) due to the WB design (i.e., having to consult the writeset).

The relative performance of the considered STMs is almost reversed when considering  workload  {\sf B} (Fig.\ref{fig:plot_throughput_bank_high_con_mram}). Here, NOrec shines, as it wastes less time processing aborted transactions (Fig.\ref{fig:plot_phases_bank_high_con_mram}) for two main reasons: i) in NOrec transactions wait until the global sequence lock is free before starting, which  acts as a contention management mechanism~\cite{Scherer2005PODC,spear2009comprehensive}; {ii)  the abort cost  is lower in NOrec since it does not need to update any ORec.


The ETL-based variants of VR, i.e., two most competitive solutions for workload {\sf A}, here stop scaling at around 4 tasklets and their peak throughput is $\sim$40\% lower than NOrec's. In fact, in this workload, a read on the a data item is always followed by a write. Hence, the use of read-write locks provides limited benefits, unlike in workload  {\sf A}  where transactions access a large number of data items by solely reading them. Also in this workload, all the ETL variants  have an edge over the corresponding CTL counterparts (for the same reasons discussed when analysing workload {\sf B}).

Regarding the Tiny variants, they achieve a slightly worse performance than the ETL variants of VR. The main reason is that the use of VR allows detecting conflicts earlier that in Tiny, which reduces the time wasted when aborting. The CTL variants of Tiny scales better than its VR counterpart, although the two reach a similar peak throughput. In fact, the {\sf VR CTLWB} suffers  a higher abort rate than the {\sf Tiny CTLWB}: with this workload, {\sf VR CTLWB} incurs spurious aborts that are triggered when transactions attempt to upgrade a rw-lock from read mode to write mode (as if two transactions conflict on more than a data item, both of them can abort if they attempt to acquire write locks on different data items concurrently), which are instead avoided by the other designs. This also reflects into a larger wasted time for {\sf VR CTLWB} (Fig.\ref{fig:plot_phases_bank_high_con_mram}).

\paragraph{Linked-List.} In both the {\sf LC} and {\sf HC} workloads of this benchmark, the best performing STM is NOrec, whose peak throughput in the {\sf LC}/{\sf HC} workload is 6\%/15\%, better resp., than the Tiny-based solutions. The VR variants are clearly the worse performing ones, as they experience a much higher abort rate. In fact, when a transaction $T$ attempts to upgrade a rw-lock from read to write mode (i.e., in order to update a list element), it is likely to encounter that lock already acquired in read-mode by a concurrent transaction $T'$. This causes $T$ to abort, even though $T'$ may also later abort. Conversely, in the invisible reads based designs, if a transaction $T$ aborts, this is always due to a conflict with a committed transaction.

As for the comparison between NOrec and the Tiny-based variants (which achieve similar performance in these workloads), the reasons underlying the NOrec gains differ depending on the considered workload: in the {\sf LC} workload, despite incurring a slightly higher abort rate, NOrec has an edge over the Tiny variants as it can process reads more efficiently, due to its simpler logic (which overall spares 2 MRAM access w.r.t. Tiny to process a read request); in the {\sf HC} workload, NOrec and the Tiny variants have a similar abort rate, but NOrec wastes less time aborting (by performing more frequent validations, NOrec  detects conflicts earlier than Tiny).

Regarding lock timing, the ETL variants of Tiny and VR have a slight advantage over their CTL counterparts, although this design choice has a less strong impact (especially for the Tiny-based approaches) than in the ArrayBench workload. This can be explained by noting that transactions have smaller writesets in this workload. Thus, the extra costs incurred by CTL when reading (i.e., scanning the writeset to determine if the item was previously written) are reduced. As for the write policy (WB vs WT), it has a negligible impact on performance (analogously to what previously observed).

\paragraph{KMeans.} In the {\sf LC} scenario(Fig.\ref{fig:plot_throughput_kmeans_14_dims_15_centers_mram}), we 
observe an almost linear scalability for NOrec and for the ETL-
based variants of Tiny and VR, which all achieve very similar peak throughput. NOrec has a slight edge over these solutions, mostly thanks to its more efficient handling of read and write operations (Fig.\ref{fig:plot_phases_kmeans_14_dims_15_centers_mram}). We can also observe that, despite the various STMs experiencing quite different abort rates (ranging from $\sim$5\% to $\sim$80\%), this does not impact significantly throughput. In KMeans LC, most of the time is spent in non-transactional code, which explains why the choice of the STM implementation has only a limited impact on performance.

This is not the case for the {\sf HC} scenario, though. As the number of centroids decreases by a factor of 7.5$\times$, the benchmark spends a much larger fraction of time executing transactional code which increases contention drastically. Thus, the performance gaps among the STMs amplify, with NOrec achieving $\sim$22\% higher throughput than the ETL variants of Tiny, which, in turn, are followed closely by the ETL variants of VR. This workload is the one that shows the largest penalty for the CTL design, for both its Tiny-based and VR-based variants. Interestingly, the CTL variant of Tiny suffers a much lower abort rate than its ETL counterparts (Fig.\ref{fig:plot_aborts_kmeans_14_dims_2_centers_mram}). In fact, despite the Tiny CTL design avoiding some spurious aborts incurred by the Tiny ETL design (by postponing lock acquisition and reducing lock duration), the CTL variants detect conflicts later (leading to more wasted work). Furthermore, even in the absence of contention, CTL implementations spend significant more time in the read and commit phases (Fig.\ref{fig:plot_phases_kmeans_14_dims_2_centers_mram}). 

\paragraph{Labyrinth.} In Labyrinth {\sf S} and {\sf L} (Fig.\ref{fig:plot_throughput_labyrinth_low_con_mram} and~\ref{fig:plot_throughput_labyrinth_high_con_mram}), all implementations achieve a similar peak throughput at $\sim$5 tasklets. Slightly larger performance gaps are observable in workload {\sf S}, whereas the performances of the various STMs are closer in workload {\sf L}. Unlike in KMeans {\sf LC}, where most of the time is spent in non-transactional code, in this benchmark  almost 100\% of the time is spent in transactional code.

The time breakdown plots show that, unlike all other considered benchmarks, in Labyrinth the time spent processing (``Other (Executing)'' in Fig.\ref{fig:plot_phases_labyrinth_128_128_3_mram}) during transactions is the dominating cost at 1 tasklet. Further, this time grows drastically at 11 tasklets (especially in workload {\sf L}), overshadowing the time spent in STM related activities. This increase is due to the characteristics of these workloads (strongly memory bound) that lead to under-utilize the DPU's pipeline, causing the DPU performance to saturate with less than 11 tasklets (\S\ref{sec:upmem_pim} and~\cite{upmemhardware}). Thus, scalability is limited not only by  contention at the STM level, but also  at the hardware level.

At the STM level, contention is  similar for all variants, except for the VR-based STMs (Fig.\ref{fig:plot_aborts_labyrinth_low_con_mram} and~\ref{fig:plot_aborts_labyrinth_high_con_mram}). However, the additional aborts in the VR-based STMs are associated with a very short transaction that is used to extract jobs from a shared queue. This transaction is more susceptible to trigger spurious aborts in the VR-based solutions and undergoes a larger number of retries. However, due to its relatively short execution, its aborts have limited impact on  performance.

\subsubsection{Main Conclusions}

Below we summarize the main conclusions of the study presented above. To aid our analysis, we report in Fig.\ref{fig:violin_plot_mram} the distribution of the peak throughput of each STM  normalized by the peak throughput of the best STM for each workload. 

\paragraph{Metadata granularity.} NOrec achieved the best (average and median) performance across all benchmarks. This can be attributed to two main factors: F1) NOrec manipulates a much smaller amount of metadata, which translates into faster read and write phases; F2) NOrec tends to experience lower abort rates in high contention scenarios since: a) NOrec wait until the global sequence lock is free before starting transactions; b) due to F1, the duration of the transaction execution phase is reduced, which helps reducing conflict probability~\cite{di-sanzo11,mascots-daniel}. 

However, due to its coarse metadata granularity, NOrec performance can be severely hampered in workloads (e.g., ArrayBench {\sf LC}) where transactions have  large readsets and a low conflict probability. In these settings, NOrec undergoes frequent, yet unnecessary, readset validations that can introduce prohibitive overheads.

\paragraph{Read visibility.} The designs based on invisible reads (i.e., Tiny-based) tend to outperform the VR-based variants in high-contention scenarios, where the latter tend to suffer frequent spurious aborts. However, in low contention workloads in which transactions issue a large number of reads, e.g., ArrayBench {\sf LC}, the best VR design variant is $\sim$2$\times$ faster than the best Tiny-based variant, given that: i) VR avoids the cost of readset validation; ii) the overhead of tracking readers (which requires acquiring a rw-lock in read mode) is quite low on the UPMEM system and is outweighed by the benefits provided by the use of VR's simpler design (that spares several MRAM read accesses when compared to the Tiny variants).

\paragraph{Lock timing.}  ETL is overall more competitive approach than CTL, at least for the considered workloads, as the main potential benefit of CTL, i.e., reducing abort rate, is largely outweighed by its drawbacks, i.e., more wasted work in case of abort and higher read cost.

\paragraph{Write policy.} The write policy (WB vs WT) has very limited impact on performance in all the considered workloads. Indeed, when the STM metadata are kept in MRAM, the instrumentation overhead is mostly dependent on the number of MRAM accesses performed by the STM and the write policy choice has a limited impact on this.

\paragraph{No one-size-fits-all solution.} Overall, our results show that no STM delivers optimal performance across all the considered workloads and that even the solution that is on average most competitive (i.e, NOrec) can be up to 2$\times$ slower when faced with non-favourable workloads.

\begin{figure}[t!]
    \centering
    \subfloat[Metadata in MRAM\label{fig:violin_plot_mram}]{
    \includegraphics[width=0.95\linewidth]{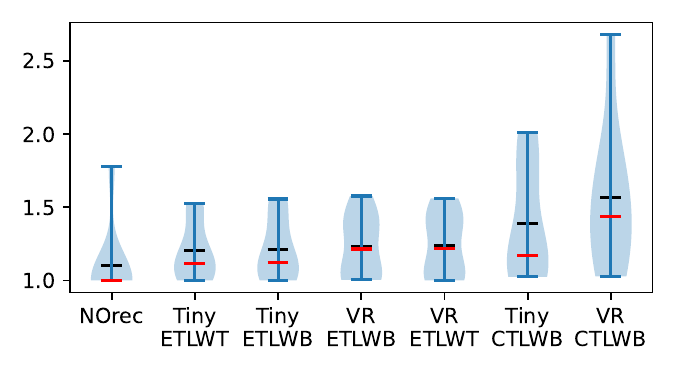}
    }\\
    \subfloat[Metadata in WRAM\label{fig:violin_plot_wram}]{
    \includegraphics[width=0.95\linewidth]{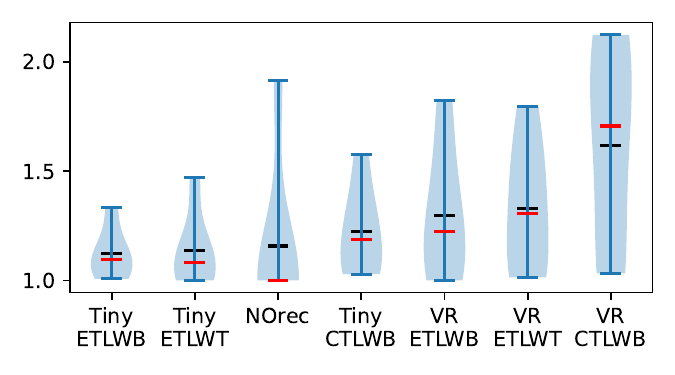}}   
    \caption{Distribution across all workloads of the ratio between the peak throughput of the best STM for a given workload and the peak throughput of a given STM for that workload. Lower is better.}
\end{figure}

\subsubsection{STM metadata hosted in WRAM} \label{sec:wram}
Due to space constraints we include the plots for the case of metadata hosted in WRAM as an appendix, but we discuss the main conclusions below.

$\bullet~$ The use of WRAM to maintain STM metadata reduces the STM instrumentation overheads significantly, with beneficial effects both on peak throughput and scalability: hosting the STM metadata in WRAM reduces transaction duration, which, in turn, reduces the likelihood of conflicts. The speed-ups achievable via the use of WRAM are strongly related to the the fraction of time  spent running transactional code. For instance, in KMeans {\sf LC}, which spends a negligible percentage of time executing transactions, the throughput gains are $\sim$5\%. In the remaining workloads, which  spend most of the time in transactional code, the gains range from 2.46$\times$ to 5.1$\times$, with a geometric mean of 2.86$\times$. Based on these results, our recommendation is to host STM metadata in MRAM if applications make little use of transactions (to reserve WRAM capacity for applications' needs). If applications heavily rely on transactions, the gains achievable by using WRAM to accelerate the STM implementation can be substantial; however, one should factor in the possible slow-down caused by reducing the available WRAM capacity for other applications' purpose.

$\bullet~$ Although NOrec remains the most competitive solution in 75\% of the workloads, the two ETL variants of Tiny become the  most competitive solutions based on the average normalized peak-throughput  (Fig.\ref{fig:violin_plot_wram}). In fact, when the Orecs are allocated in WRAM, their access latency is reduced, which improves the efficiency of Orecs-based solution.

$\bullet~$ The choice of the write policy (WB vs WT) has stronger impact than when hosting the STM metadata in MRAM, as , by accelerating the access to the STM metadata via WRAM, the relative cost of applying the transaction's write to MRAM is amplified. As expected, WB is favoured in high contention workloads (up to 14\% throughput increase in ArrayBench {\sf B}) and WT in low contention ones (up to $\sim$5\% throughput increase in Linked-List {\sf LC}).



\begin{figure*}[t!]
    \subfloat[Speedup for KMeans.\label{fig:throughput_kmeans_dist}]{
    \includegraphics[width=0.49\linewidth]{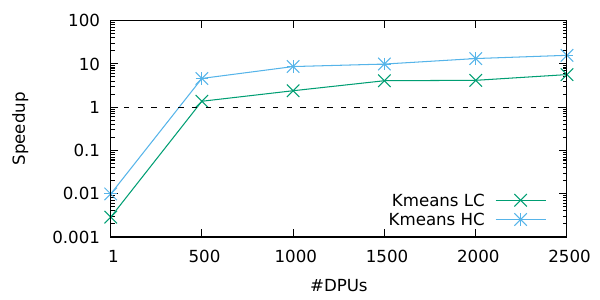}
    }
    \hfill
    \subfloat[Speedup for Labyrinth.\label{fig:plot_throughput_labyrinth_multi_dpu}]{
    \includegraphics[width=0.49\linewidth]{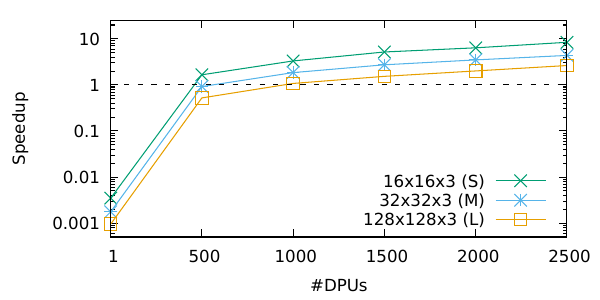}
    }
    \caption{Speedup with respect to CPU-based implementations of KMeans and Labyrinth.}
\end{figure*}

\begin{figure}[t!]
    \centering
    \includegraphics[width=0.96\linewidth]{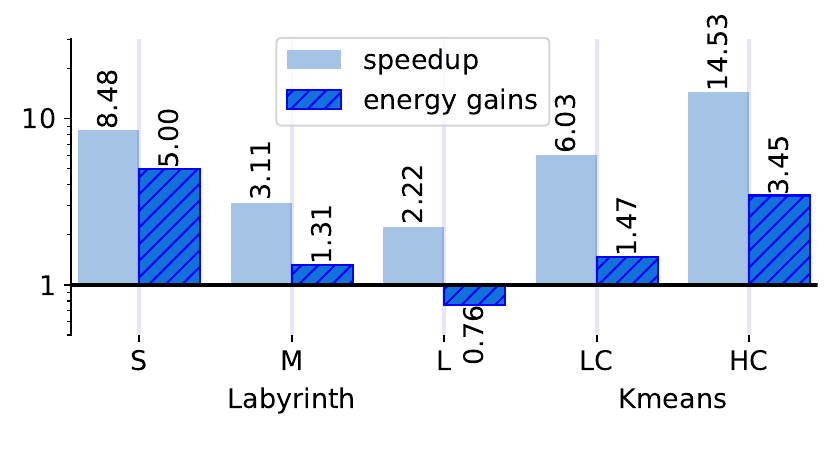}
    \caption{Speedup and energy gains for 2500 DPUs with respect to CPU-based implementations of KMeans and Labyrinth.}
    \label{fig:energy_plot}
\end{figure}

\subsection{Performance and energy gains with respect to CPU} \label{sec:multi-dpu}

This section assesses the performance and energy gains achievable by using PIM-STM to accelerate STM-based applications originally developed for CPUs. To this end, we ported KMeans and Labyrinth to operate  on the thousands of  DPUs provided by the current UPMEM system. We first  describe our multi-DPU porting of KMeans and Labyrinth (\S\ref{sec:multi-dpu-bench}). Next, we analyse performance (\S\ref{sec:multi-dpu-performance}) and energy (\S\ref{sec:multi-dpu-energy}) gains.

\subsubsection{Multi-DPU Benchmarks}
\label{sec:multi-dpu-bench}

We use our multi-DPU porting of Kmeans and Labyrinth to illustrate two  approaches that take advantage of inter-dpu parallelism: i) having the different DPUs cooperate to solve a single problem (Kmeans), or ii) letting each DPU solve an independent instance of the same problem (Labyrinth).

\paragraph{Kmeans.} We adapted KMeans to have the CPU  distribute disjoint shards of the input  points to each DPU. The DPUs operate in parallel on a private copy of the centroids. At the end of each round, the DPUs communicate their locally updated centroids to the CPU, which merges  these updates and communicates the new centroids to the DPUs to start a new round. For the sake of fairness, we configure both the CPU and DPU implementations to perform the same number  (3) of rounds. Also, in this experiment each DPU is assigned 200K input points. Thus, as we vary the number of used DPUs, we also vary the total number of inputs points both  for the DPU and  CPU implementations.

\paragraph{Labyrinth.} In our multi-DPU porting of Labyrinth, the CPU schedules the execution of independent instances of circuit routing problems on different DPUs; the CPU initiates a job by transferring the problem inputs to the DPUs; the DPUs use the PIM-STM library to solve their own problem instance and report back the updated grid to the CPU.

\paragraph{General considerations.} On the DPU-side, we used for both benchmarks the NOrec STM implementations, which we configured to use the number of tasklets that provide peak throughput. On the CPU-side we use the NOrec implementation, which we also we also configured to utilize the optimum number of threads, namely 4/8 threads for KMeans/Labyrinth, resp. As our  Labyrinth porting is used to solve independent problems,  on the CPU side we execute 4 independent processes in parallel (each using 8 threads). This is done to ensure the full utilization of all the 32 CPU {hardware} threads. Finally, for KMeans we allocate the STM metadata in WRAM; this is not possible for Labyrinth (as WRAM has insufficient capacity to maintain transactions' readsets and writesets), so we allocate the STM metadata in MRAM.

\subsubsection{Performance gains}
\label{sec:multi-dpu-performance}




Fig.\ref{fig:throughput_kmeans_dist} reports the speedups for KMeans {\sf LC} and {\sf HC} with respect to the CPU-based implementation, as we increase the number of DPUs. The performance of a single DPU is $\sim$100$\times$/$\sim$300$\times$ slower than the CPU for {\sf LC}/{\sf HC}, resp.  As we increase the number of DPUs, we observe performance gains for the PIM-STM-based implementation starting at around 300/400 DPUs that grow linearly up to approx.~14$\times$/6$\times$ for  {\sf HC}/{\sf LC}, resp.. In fact, as the number of DPUs/input size increases, the execution time of the DPU-based version remains approx. constant, whereas it grows linearly for the CPU-based implementation.

Labyrinth (Fig.\ref{fig:plot_throughput_labyrinth_multi_dpu}) shows a similar trend: the CPU strongly outperforms a single DPU, but as the number of DPUs increases, the speed-ups of the PIM-based implementation grow linearly. In fact, the  throughput for the PIM-based implementation grows linearly with the number of DPUs, whereas it remains constant for the CPU. The peak gains  at 2500 DPUs range from 8.48$\times$ (smallest grid size) to 2.22$\times$ (largest grid size). As discussed in \S\ref{sec:mram}, as the grid size increases, the workload characteristics become less favourable for the UPMEM system, causing its pipeline to saturate at about half of its capacity and limiting the effective intra-DPU parallelism. This explains why its competitiveness w.r.t.~the CPU-based implementation decreases as the input size grows.

\subsubsection{Energy efficiency}\label{sec:multi-dpu-energy}
 Fig.\ref{fig:energy_plot} presents the speedup and energy gain when using all DPUs for all the workloads of our multi-DPU benchmarks. Energy gain is computed as the ratio of the energy used by the CPU and by the DPU. Given the lack of energy counters on the UPMEM system, we estimate the energy consumed by UPMEM for a given workload as its thermal design power (TDP), namely 370W~\cite{Falevoz2023pecs} when using all DPUs,  multiplied by the workload's execution time. Conversely, for the CPU-based implementations,  we  measure the energy consumed both for the CPU and memory sub-systems via the RAPL~\cite{david2010rapl} library. 


In Fig.\ref{fig:energy_plot} we observe lower energy gains than performance gains, and even a 31.5\% higher energy consumption for Labyrinth \textsf{L} (whose speedup is 2.22$\times$). Overall, these results show that the current UPMEM system has an excellent performance potential, but is not equally competitive in terms of energy efficiency. Fortunately,  this gap is expected to be narrowed in the next generation of the UPMEM system, which is expected to be 35\% more energy efficient~\cite{Falevoz2023pecs}. 
%



\section{Conclusions and future work}


This work tackled the problem of how to develop efficient STM implementations for PIM, by introducing PIM-STM, a library that provides a range of alternative STM implementations for UPMEM (the first commercial PIM system). 

Via an extensive experimental study, we investigated the efficiency of alternative STM designs as well as quantified the impact of using different memory tiers provided by the UPMEM system to maintain the STM metadata. We also assessed the performance and energy gains achievable by using PIM-STM to accelerate two popular TM benchmarks originally designed for traditional CPU-based systems.

The introduction of the PIM-STM library  paved the way for future work aimed at further evaluating the effectiveness and efficiency of the STM abstraction in a broader range of domains. A relevant domain, where STM is already being employed~\cite{dickerson2017adding,gelashvili2023block}, is parallelization of block-chains, namely to accelerate both the mining and validation of new blocks. Another interesting research question is how to leverage the PIM-STM library in order to implement non-transactional concurrent data-structures (such as linked list or hashmaps) that can be distributed across multiple DPUs, so as to exceed the memory capacity of a single DPU. Since PIM-STM transparently regulates concurrency within the boundaries of individual DPUs, the key problems that remain to be investigated in order to pursue this goal are i) how to distribute operations across DPUs efficiently, and ii) how to coordinate operations' execution whenever these require updating atomically the state of multiple DPUs.

\section*{Acknowledgments}                            
  The authors would like to thank Professor Ryan Stutsman for his valuable feedback during the shepherding process as well as   the anonymous reviewers for their insightful comments. We are also grateful to UPMEM  for having provided us with free access to their innovative platform and for their constant support and guidance about the usage and optimization of their system. This work was supported by national funds through FCT, Fundação para a Ciência e a Tecnologia, under project UIDB/50021/2020 (DOI:10.54499/UIDB/50021/2020) and via IAPMEI under project C645008882\mbox{-}00000055.PRR (NextGenerationEU/EU Recovery and Resilience Plan).

\bibliographystyle{plain}
\bibliography{library}

\newpage
\appendix
\section{STM metadata hosted in WRAM}
This appendix contains the results obtained for the scenario in which the STM metadata is hosted in WRAM --- which we could not include in the main body of the paper for space constraints. The corresponding plots are reported in Figures~\ref{fig:wram1} and~\ref{fig:wram2}.

Note that these results do not include Labyrinth, as the transactions' readsets and writesets for this benchmark (when using 11 tasklets) exceed the total WRAM capacity. Also, in the workload  ArrayBench \textsf{A}, the lock table used by the ORec-based STM designs, namely Tiny and VR, exceeds the WRAM. So, in that workload, we configure these STMs to allocate the lock-table in MRAM (and all other metadata in WRAM). This capacity limitation is expected to favour, at least in this workload, NOrec --- which, does not rely on a lock table, can allocate all of its meta-data (sequence lock and read/writesets in WRAM).

\begin{figure*}[ht]
    \subfloat[Throughput ArrayBench A\label{fig:plot_throughput_bank_10000_acc}\vspace{-5pt}]{
        \includegraphics[width=0.24\linewidth]{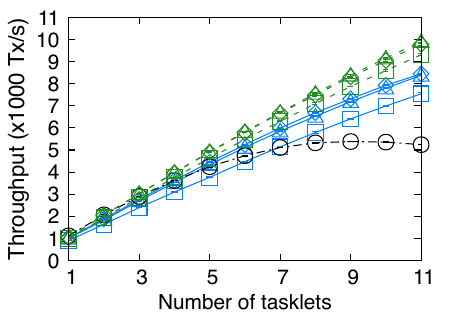}
    }
    \subfloat[Throughput ArrayBench B\label{fig:plot_throughput_bank_high_con}\vspace{-5pt}]{
        \includegraphics[width=0.24\linewidth]{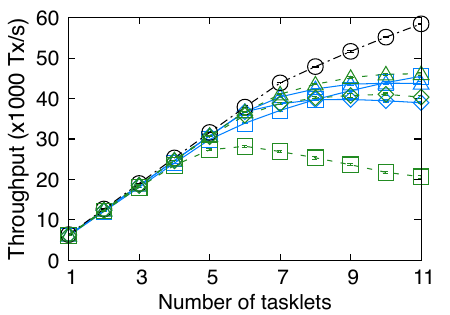}
    }
    \subfloat[Throughput linked-list LC\label{fig:plot_throughput_linked_list_10_upd}\vspace{-5pt}]{
        \includegraphics[width=0.24\linewidth]{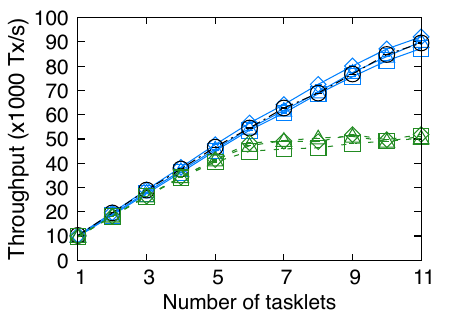}   
    }
    \subfloat[Throughput linked-list HC\label{fig:plot_throughput_linked_list_50_upd}\vspace{-5pt}]{
        \includegraphics[width=0.24\linewidth]{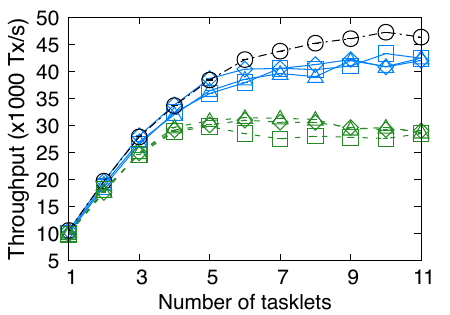}
    }
    \hfill
    \medskip
    \subfloat[Abort rate ArrayBench LC\label{fig:plot_aborts_bank_10000_acc}\vspace{-5pt}]{
        \includegraphics[width=0.24\linewidth]{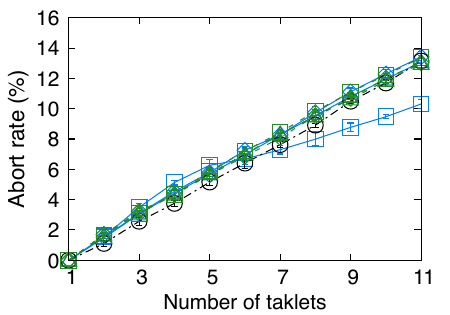}
    }
    \subfloat[Abort rate ArrayBench HC\label{fig:plot_aborts_bank_high_con}\vspace{-5pt}]{
        \includegraphics[width=0.24\linewidth]{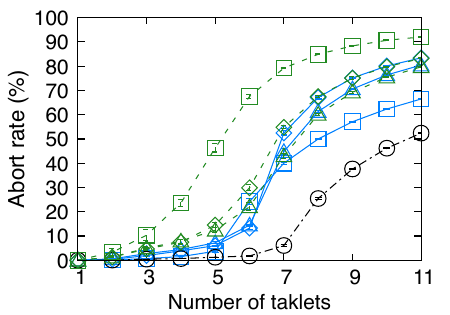}
    }    
    \subfloat[Abort rate linked-list LC\label{fig:plot_aborts_linked_list_10_upd}\vspace{-5pt}]{
        \includegraphics[width=0.24\linewidth]{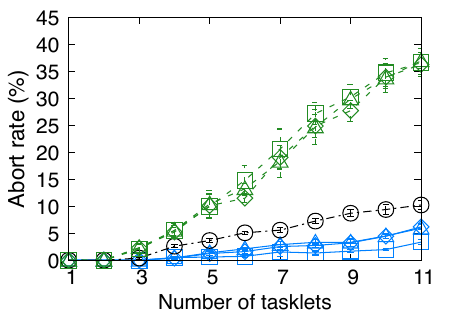}
    }
    \subfloat[Abort rate linked-list HC\label{fig:plot_aborts_linked_list_50_upd}\vspace{-5pt}]{
        \includegraphics[width=0.24\linewidth]{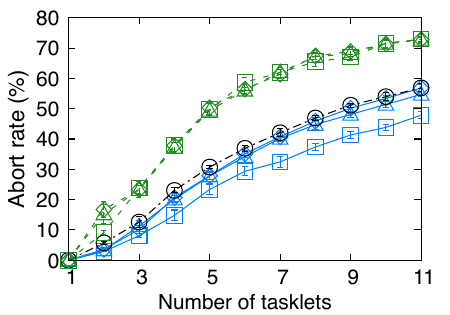}
    }
    \hfill
    \medskip   
    \subfloat[Phases ArrayBench LC\label{fig:plot_phases_bank_bank_10000_acc}\vspace{-5pt}]{
        \includegraphics[width=0.24\linewidth]{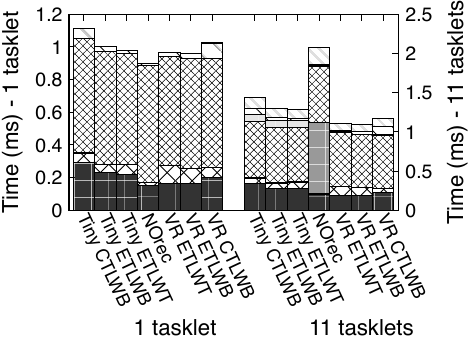}
    }
    \subfloat[Phases ArrayBench HC\label{fig:plot_phases_bank_high_con}\vspace{-5pt}]{
        \includegraphics[width=0.24\linewidth]{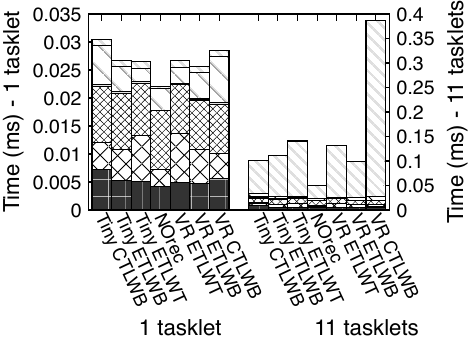}
    }
    \subfloat[Phases linked-list LC\label{fig:plot_phases_linked_list_10_upd}\vspace{-5pt}]{
        \includegraphics[width=0.24\linewidth]{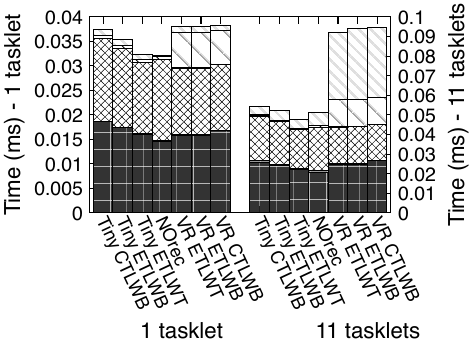}
    }
    \subfloat[Phases linked-list HC\label{fig:plot_phases_linked_list_50_upd}\vspace{-5pt}]{
        \includegraphics[width=0.24\linewidth]{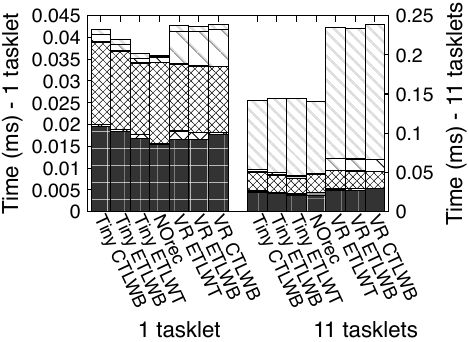}
    }
    \medskip

    \hfill
    \begin{tabular}{c c c c}
         \includegraphics[width=0.13\linewidth]{Graphs/legend_1.pdf} & \includegraphics[width=0.12\linewidth]{Graphs/legend_2.pdf} & $\qquad\qquad\qquad$\includegraphics[width=0.19\linewidth]{Graphs/legend_phases_1.pdf} & \includegraphics[width=0.17\linewidth]{Graphs/legend_phases_2.pdf} \\
    \end{tabular}
    \hfill
    \hfill

    \vspace{-4pt}
    \caption{Throughput, abort rate and time breakdown for the ArrayBench and linked-list benchmarks with metadata in WRAM}\label{fig:wram1}
\end{figure*}

We analyse each benchmark in the following.

\paragraph{ArrayBench.}
Focusing on Tiny's encounter time locking (ETL) implementations ({\sf Tiny ETLWB} and {\sf Tiny ETLWT}) and analysing Fig.\ref{fig:plot_throughput_bank_10000_acc} and~\ref{fig:plot_throughput_bank_high_con}, we can observe that write back (WB) and write through (WT), yield similar performance in the ArrayBench {\sf A} scenario. In the ArrayBench {\sf B} scenario, WB achieves 4\% higher throughput. This is expected, as  WB is more efficient in scenarios where there are a lot of aborts. We note that the performance gap between the WB and WT policy is amplified w.r.t.~MRAM case. This can be explained by considering that, when the STM metadata is kept in WRAM, the cost for accessing them decreases significantly. Consequently, the relative gains of avoiding MRAM accesses to undo the writes of aborted transactions in the high contention scenario (enabled by the WB policy) grows.

Comparing Tiny's write back (WB) based implementations ({\sf Tiny CTLWB} and {\sf Tiny ETLWB}) and analysing Fig.\ref{fig:plot_throughput_bank_10000_acc}, ETL yields 2\% better performance in the ArrayBench {\sf A} scenario. Taking into consideration Fig.\ref{fig:plot_phases_bank_bank_10000_acc}, we can see that this difference in performance, stems form a longer read and commit phases in the CTL implementation. We argue that this difference arises due to the CTL implementation needing to check the entire write set (for previous writes to the same position) every time a read is performed. Additionally, CTLWB needs to perform more work at commit time (i.e., acquire locks and write values to memory) resulting in a longer commit phase. On the other hand, in ArrayBench {\sf B} scenario (Fig.\ref{fig:plot_throughput_bank_high_con} and~\ref{fig:plot_phases_bank_high_con}), CTLWB yields around 2\% higher performance than ETLWB. This can be explained by the 10\% higher abort rate incurred by the ETL implementation. The increased abort rate arises due to ETL maintaining locks during the entire execution of a transaction which causes spurious aborts that can be avoided with the CTL design.

From Fig.\ref{fig:plot_throughput_bank_10000_acc} we can also observe that NOrec has the lowest throughput in the ArrayBench {\sf A} scenario, as we had already observed in the scenario of STM metadata hosted in MRAM. As already discussed, this can be explained by the fact that NOrec needs to revalidate the entire read set every time a concurrent transactions commits. As the number of tasklets increases, so does the amount of validation. In the ArrayBench {\sf B} scenario (Fig.\ref{fig:plot_throughput_bank_high_con}), NOrec outperforms the best Tiny and VR variant by 20\%. This difference in throughput, can be attributed to two factors. First, NOrec has a considerably lower abort rate (Fig.\ref{fig:plot_aborts_bank_high_con}). Second, since NOrec does not use ownership records, it does not incur the overhead associated with maintaining such records, which results in a shorter write phase (Fig.\ref{fig:plot_phases_bank_high_con}).

Comparing the WB and WT variants of the VR-ETL STM and analysing Fig.\ref{fig:plot_throughput_bank_10000_acc} and~\ref{fig:plot_throughput_bank_high_con}, we  observe that the two write policies yield similar performance in the ArrayBench {\sf A} scenario. In the ArrayBench {\sf B} scenario, WB achieves 14\% higher throughput. This is the case because WB, is more efficient in when dealing with high abort rates. WB writes new values in a log instead of writing to memory. Thus, when aborting, a TM implementation that uses WB (instead of WT) does not incur the overhead of restoring (or undo) writes. Also, in this case, we observe an increase in the relative impact on performance of the writing policy (for the same reasons discussed above).

As for the lock timing policy,
ETL yields 4\% better performance in the ArrayBench {\sf A} scenario. Taking into consideration Fig.\ref{fig:plot_phases_bank_bank_10000_acc}, this difference in performance, stems form 2 factors: longer read and commit phases in the CTL implementation. We argue that the read phase is longer in the CTL implementation due to needing to check the entire write set (for previous writes to the same position) every time a read is performed. Additionally, \textsf{VR CTLWB} needs to perform more work at commit time (i.e., acquire locks and write values to memory) resulting in a longer commit phase. In the ArrayBench {\sf B} scenario (Fig.\ref{fig:plot_throughput_bank_high_con} and~\ref{fig:plot_phases_bank_high_con}), \textsf{VR ETLWB} yields around 64\% higher performance than \textsf{VR CTLWB}. This can be explained by considering that \textsf{VR CTLWB} suffers a significantly higher abort rate in, which  arises because multiple concurrent transactions try to acquire write locks roughly at the same time (commit phase).

\paragraph{Linked-List.} In the LC scenario with 10\% update operations (add, remove), \textsf{Tiny ETLWT} performs 5\% better than the other Tiny variants, mainly due to a shorter read phase (Fig.\ref{fig:plot_phases_linked_list_10_upd}). This is the case because \textsf{Tiny ETLWT} employs WT (i.e. the writes are performed directly into memory, instead of being buffered), hence reads do not need to check the write set for previously buffered writes. Similarly, the {\sf Tiny ETLWB} implementation performs better than \textsf{Tiny CTLWB} for the same reason, a shorter read phase due to {\sf Tiny ETLWB} needing to perform less validation when reading. In the high contention scenario, presented in Fig.\ref{fig:plot_throughput_linked_list_50_upd}, {\sf Tiny ETLWB} performs similarly to \textsf{Tiny ETLWT}. \textsf{Tiny CTLWB} has the highest performance out of the Tiny implementations, due to a 8\% lower abort rate.

In the low contention scenario (Fig.\ref{fig:plot_throughput_linked_list_10_upd}), NOrec has slightly lower performance than WBET. In fact, when the STM metadata is stored in WRAM, the performance penalty for accessing the ORecs is reduced, which benefits both VR and, in particular, Tiny. However, in the HC scenario (Fig.\ref{fig:plot_throughput_linked_list_50_upd}), NOrec performs 9\% better than the best Tiny implementations. We argue that this difference in throughput arises due to NOrec spending less time performing wasted work.

The VR implementations have the lowest throughput in both the LC scenario (86\% lower than ETLWT) and HC scenarios (58\% lower than NOrec) (Fig.\ref{fig:plot_throughput_linked_list_10_upd} and~\ref{fig:plot_throughput_linked_list_50_upd}). This is the result of a significantly higher abort rate (Fig.\ref{fig:plot_aborts_linked_list_10_upd} and Fig.\ref{fig:plot_aborts_linked_list_50_upd}). 

\paragraph{Kmeans.} In the LC scenario, Fig.\ref{fig:plot_throughput_kmeans_14_dims_15_centers} all the TM implementations perform similarly, as we had already observed in the scenario of STM metadata hosted in MRAM (and for the same reasons therein discussed).

In the HC scenario, Fig.\ref{fig:plot_throughput_kmeans_14_dims_2_centers}, NOrec has the best performance, although the performance gap with respect to the ETL-based variants of the Orec based STMs (Tiny and VR) is significantly reduced. Also in this case, the choice of storing the STM metadata in MRAM ends up benefiting the Orec  based approaches, as we had already observed in the linked list HC workload.

In general the VR implementations perform worse than their Tiny counterparts (e.g., \textsf{VR ETLWT} performs worse than \textsf{Tiny ETLWT}). Also in this scenario, this is due to the fact VR implementations suffer of higher abort rates (Fig.\ref{fig:plot_phases_kmeans_14_dims_2_centers}).

Despite having a relatively low abort rate (when compared with the other implementations), CTLWB has very low throughput (Fig.\ref{fig:plot_throughput_kmeans_14_dims_2_centers} and~\ref{fig:plot_aborts_kmeans_14_dims_2_centers}). We had already observed a similar phenomenon when considering the case of STM metadata hosted in MRAM and the reasons underlying it are the same: in CTL the gains stemming from reducing abort rate are outweighed by wasting more work when transactions do abort. This is less noticeable in ArrayBench and Linked-List because the transactions of these benchmarks are relatively smaller. Since KMeans has longer transactions, the wasted work is noticeable in the throughput. In fact, CTLWB on average spends 10$\times$ more time performing transactions that ultimately abort than the other TM implementations.

\begin{figure}[h!t]
    \subfloat[Throughput in KMeans LC\label{fig:plot_throughput_kmeans_14_dims_15_centers}\vspace{-5pt}]{
        \includegraphics[width=0.46\linewidth]{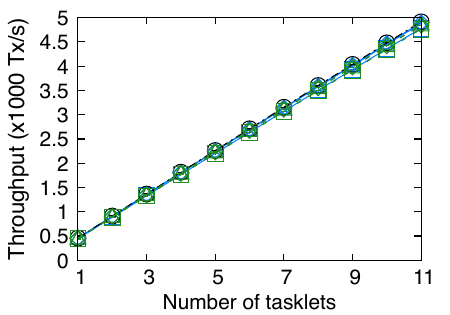}
    }
    \subfloat[Throughput in KMeans HC\label{fig:plot_throughput_kmeans_14_dims_2_centers}\vspace{-5pt}]{
        \includegraphics[width=0.46\linewidth]{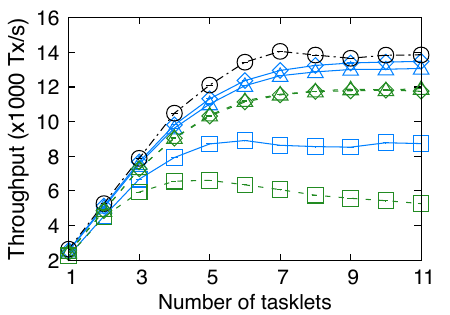}
    }
    \hfill
    \medskip
    \subfloat[Abort rate in KMeans LC\label{fig:plot_aborts_kmeans_14_dims_15_centers}\vspace{-5pt}]{
        \includegraphics[width=0.46\linewidth]{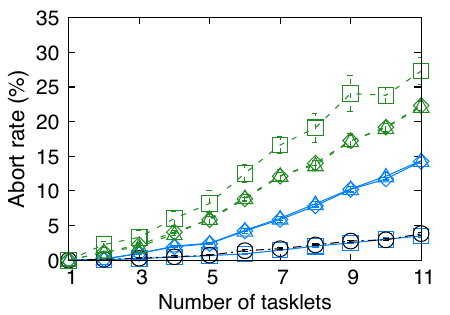}
    }
    \subfloat[Abort rate in KMeans HC\label{fig:plot_aborts_kmeans_14_dims_2_centers}\vspace{-5pt}]{
        \includegraphics[width=0.46\linewidth]{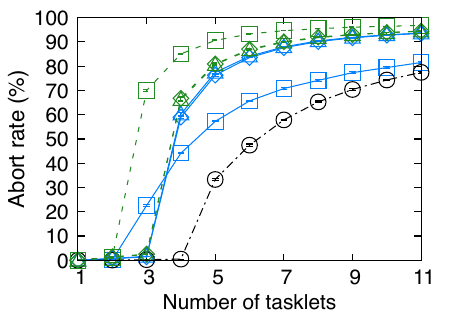}
    }
    \hfill
    \medskip   
    \subfloat[Phases in KMeans LC\label{fig:plot_phases_kmeans_14_dims_15_centers}\vspace{-5pt}]{
        \includegraphics[width=0.48\linewidth]{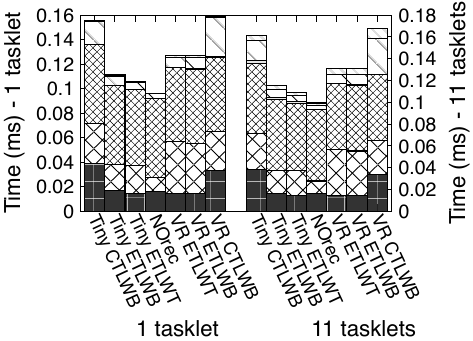}
    }
    \subfloat[Phases in KMeans HC\label{fig:plot_phases_kmeans_14_dims_2_centers}\vspace{-5pt}]{
        \includegraphics[width=0.48\linewidth]{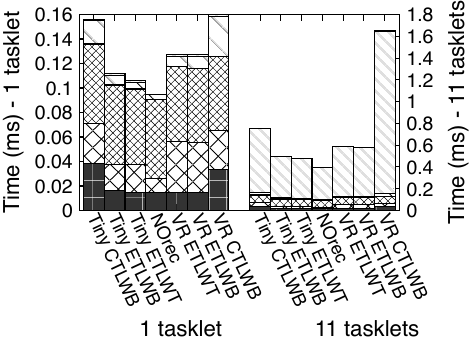}
    }
    \hfill
    \medskip
    $\qquad$
    \begin{tabular}{c c}
    \centering
         \includegraphics[width=0.28\linewidth]{Graphs/legend_1.pdf} & $\qquad$\includegraphics[width=0.41\linewidth]{Graphs/legend_phases_1.pdf} \\[-4pt]
         $\,\,\,\,\,$\includegraphics[width=0.26\linewidth]{Graphs/legend_2.pdf} & $\qquad\,\,\,\,\,$\includegraphics[width=0.38\linewidth]{Graphs/legend_phases_2.pdf}
    \end{tabular}

    \vspace{-16pt}
    \caption{Throughput, abort rate and time breakdown for the KMeans benchmark with metadata in WRAM}\label{fig:wram2}
\end{figure}

\end{document}